\documentclass[lettersize,journal]{IEEEtran}
\usepackage{amsmath,amsfonts}
\usepackage{algorithmic}
\usepackage{algorithm}
\usepackage{array}
\usepackage{multirow}
\usepackage[caption=false,font=normalsize,labelfont=sf,textfont=sf]{subfig}
\usepackage{textcomp}
\usepackage{stfloats}
\usepackage{url}
\usepackage{verbatim}
\usepackage{graphicx}
\usepackage{hyperref}
\usepackage{cite}
\begin{document}

\title{Arctic-SAR: A Cross-Sensor RADARSAT Constellation Mission and Sentinel-1 Dataset and Fusion Benchmark for Multi-Task Sea Ice Mapping}

\author{Niloofar~Azad,
Kyle~Gao,~\IEEEmembership{Member,~IEEE,}
Javier~Noa~Turnes,~\IEEEmembership{Student Member,~IEEE,}
Youssef~Nafea,
Linlin~Xu,~\IEEEmembership{Member,~IEEE,}
David~A.~Clausi,~\IEEEmembership{Senior Member,~IEEE}%

        % <-this % stops a space
% <-this % stops a space
\thanks{Niloofar Azad, Javier Noa Turnes, Youssef Nafea, and D. A. Clausi are with the Department of Systems Design Engineering, University of Waterloo, Waterloo, ON, Canada (e-mail: n2azad@uwaterloo.ca; jnoaturnes@uwaterloo.ca; ynafea@uwaterloo.ca;
dclausi@uwaterloo.ca).}
\thanks{Linlin Xu is with the Department of Geomatics Engineering, University of Calgary, Calgary, AB, Canada (e-mail: lincoln.xu@ucalgary.ca).}
\thanks{Kyle Gao is with the Department of Built Environment, Aalto University, Espoo, Finland (e-mail: kyle.gao@aalto.fi).}
\thanks{(Corresponding author: Niloofar Azad.)}}

% \markboth{IEEE Journal of Selected Topics in Applied Earth Observations and Remote Sensing}%
% {Azad \MakeLowercase{\textit{et al.}}: Arctic-SAR}


\maketitle

\begin{abstract}
Current sea ice mapping benchmarks are mostly built around single-sensor synthetic aperture radar (SAR) products, limiting the evaluation of cross-sensor variability and the effectiveness of combining observations from different SAR platforms. Models trained on such datasets may be sensitive to sensor-specific artifacts, radiometric characteristics, and acquisition conditions, which can reduce robustness and limit generalization. To address these limitations, this study introduces Arctic-SAR, a cross-sensor SAR dataset for Arctic sea ice mapping, constructed by matching RADARSAT Constellation Mission (RCM) dual-polarization imagery with Sentinel-1 scenes from the AI4Arctic Sea Ice Dataset. The dataset consists of 221 co-registered scenes and supports multi-task prediction of sea ice concentration, stage of development, and floe size. In addition to the dataset, we establish a benchmark for cross-sensor SAR sea ice mapping by quantifying the contribution of RCM imagery and evaluating single-sensor baselines alongside input-, feature-, and decision-level fusion strategies within a consistent multi-task U-Net framework. Using five-fold scene-level cross-validation with three independent runs per fold, we find that Sentinel-1 provides a stronger single-sensor baseline than RCM, achieving a mean combined score of $85.48\pm2.00\%$ compared with $82.56\pm3.38\%$. Input-level fusion achieves the highest combined score of $85.81\pm2.21\%$, only 0.33 percentage points higher than Sentinel-1 alone. These results suggest that RCM provides limited additional predictive information for the evaluated tasks. Although Sentinel-1 and RCM provide task- and class-dependent information, simple joint cross-sensor fusion does not consistently improve generalization over a strong single-sensor baseline. Arctic-SAR therefore provides a benchmark for cross-sensor SAR fusion research in automated Arctic sea ice mapping.
\end{abstract}


\section{Introduction}
\IEEEPARstart{A}{utomated} sea ice mapping is essential for monitoring changes in polar regions. As a key component of the cryosphere, Arctic sea ice exhibits substantial spatial and temporal variability and has direct implications for climate dynamics, marine ecosystems, and safe maritime navigation, including supporting shipping routes, fishing, resource extraction, and other maritime operations \cite{Chen2024, Lyu2022}. Therefore, accurate and efficient Arctic sea ice monitoring is crucial for enhancing our understanding of sea ice changes and long-term trends \cite{Swart2015, Ye2023, Ma2026GLFFuse}.

Satellite remote sensing is an important approach for sea ice monitoring and mapping, enabling consistent, large-scale, and frequent observations of polar regions and supporting the rapid acquisition of the spatiotemporal characteristics of sea ice \cite{Ma2026GLFFuse, Zakhvatkina2019}. As an active microwave imaging radar, SAR has become the primary sensing modality for operational sea ice mapping, as it provides all-day and all-weather imaging capabilities and supports multi-polarization observations \cite{Turnes2025ITT}. These properties enable the extraction of essential information on material thickness, surface roughness, and textural patterns \cite{Dierking2013, Soh1999}, making SAR imagery highly effective for sea ice classification and segmentation \cite{Chen2024, Boulze2020, Han2021}. However, SAR imagery is affected by speckle noise and incidence-angle-dependent variations in backscatter caused by the side-looking imaging geometry of the satellite \cite{Turnes2025ITT}. These effects can produce nonstationary image patterns \cite{Leigh2014}, especially over surfaces with specular properties. In addition, limited signal penetration into the ice, spatial variations in snow cover, and large interclass variability in physical and backscattering properties, particularly during melt and freeze-up periods, introduce further uncertainty when interpreting sea ice scenes \cite{Dierking2013, Gerland2019}. 

Recent advances in deep learning have demonstrated promising results in remote sensing prediction tasks, including SAR-based sea ice mapping, due to their powerful feature representation capabilities \cite{WangLi2021, deLoe2025, Kruk2020, Khaleghian2021}. Traditional machine learning methods for SAR-based sea ice mapping, such as Bayesian estimation \cite{Zakhvatkina2013}, support vector machines (SVMs) \cite{Leigh2014}, and artificial neural networks (ANNs) \cite{Ressel2016}, often depend on manually designed or shallow features and knowledge-driven preprocessing strategies, including feature extraction and unsupervised image segmentation \cite{Deng2005}. In contrast, deep learning models can automatically learn hierarchical representations from complex imagery, capturing both low-level textural patterns and high-level semantic information. In particular, encoder--decoder convolutional architectures have shown strong performance for the classification and segmentation of SAR imagery \cite{Rogers2024, deLoe2025, Radhakrishnan2021, Stokholm2022, Jamali2024}.

Despite these advances, many remote sensing benchmarks remain centered on single-sensor data, which can limit the evaluation of model robustness under the high complexity and variability of polar environments \cite{Stokholm2022, Boulze2020, Jamali2024, Bonafilia2020}. This is due to the challenges involved in constructing cross-sensor datasets, which require harmonizing heterogeneous observations with different spatial resolutions, backscatter characteristics, temporal information, viewing geometries, co-registration accuracies, and measured physical properties, making joint annotation and model training substantially more complex \cite{Ghamisi2019}.

Given the spatial complexity of Arctic sea ice, SAR imagery characteristics can vary substantially across sensors due to differences in polarization configuration, incidence angle distribution, imaging mode, radiometric calibration, noise characteristics, and spatial resolution \cite{Dierking2013}. As a result, deep learning models trained on a single sensor may become sensitive to sensor-specific radiometric properties, viewing geometries, and acquisition conditions, reducing robustness and limiting generalization across different sensor configurations, acquisition conditions, or environmental settings. To address the limitations of single-sensor SAR observations, previous studies have investigated the integration of SAR imagery with complementary non-SAR modalities, including multispectral and passive microwave data, as a promising strategy for combining complementary information from different sensors and platforms \cite{Han2021, Khachatrian2023}. Although these studies demonstrate the benefits of multimodal data fusion, high-quality and representative sea ice mapping benchmarks specifically designed for cross-sensor SAR fusion remain limited, and only a small number of studies have examined fusion across multiple SAR sources \cite{Gao2025, Johansson2020, Howell2018}. Cross-sensor SAR fusion can potentially exploit complementary SAR measurements acquired by different satellite systems, providing multiple backscatter realizations of similar sea ice conditions. This cross-platform variability also exposes learning models to sensor-dependent differences within the SAR domain, which can support the learning of more sensor-agnostic and transferable representations across different sensor characteristics \cite{Marsocci2025}.

Motivated by these limitations, we introduce Arctic-SAR, a new co-registered cross-sensor SAR dataset designed to advance automated Arctic sea ice mapping and support future research on cross-sensor SAR analysis. Arctic-SAR pairs RCM HH/HV imagery with Sentinel-1 observations within the AI4Arctic Sea Ice framework \cite{BuusHinkler2022}. Using this dataset, we establish a systematic multi-task benchmark to quantify the contribution of RCM imagery and evaluate the effectiveness of cross-sensor SAR fusion for automated sea ice mapping through three representative fusion strategies: input-level fusion, feature-level fusion, and decision-level fusion.

The specific contributions of this study are as follows:
\begin{enumerate}
\renewcommand{\labelenumi}{\arabic{enumi})}

\item We introduce Arctic-SAR, a new cross-sensor SAR dataset in which RCM dual-polarization imagery is spatially and temporally aligned with Sentinel-1 scenes included in the AI4Arctic Sea Ice Dataset. The Arctic-SAR dataset contains 221 scenes and supports multi-task prediction of sea ice concentration (SIC), stage of development (SOD), and floe size (FLOE) using ice-chart-derived labels.

\item We provide a detailed characterization of the proposed dataset, including analysis of the dataset composition, geographic coverage, label distributions, cross-sensor relationships between Sentinel-1 and RCM observations, and scene-level statistical analysis. These analyses describe the structure and variability of the dataset and support its value as a reference dataset for future cross-sensor SAR sea ice mapping research.

\item We establish a systematic benchmark for cross-sensor SAR fusion by comparing Sentinel-1-only and RCM-only baselines with input-level, feature-level, and decision-level fusion approaches using five-fold scene-level cross-validation with three independent training runs per fold. The results quantify the task-dependent contribution of RCM imagery, evaluate how the fusion stage affects generalization, and provide reference performance for future research on adaptive cross-sensor SAR fusion for automated sea ice mapping.
\end{enumerate}
To the best of our knowledge, Arctic-SAR is among the first cross-sensor SAR datasets specifically developed for automated Arctic sea ice mapping, providing spatially and temporally aligned RCM and Sentinel-1 observations. It is also the first benchmark to systematically evaluate input-level, feature-level, and decision-level cross-sensor SAR fusion for joint prediction of sea ice concentration (SIC), stage of development (SOD), and floe size (FLOE).

The dataset construction pipeline and benchmark implementation are publicly available online.\footnote{
Dataset construction pipeline:
\url{https://github.com/RS-VIP/Arctic-SAR_RCM_Sentinel1_Dataset_Construction_Pipeline}.
\par
Fusion benchmark:
\url{https://github.com/NiloofarAZAD/Arctic-SAR-Fusion_Benchmark}.
}

The remainder of this article is organized as follows. 
\hyperref[sec:related_work]{Section~II} reviews single-sensor and cross-sensor deep learning approaches for SAR-based sea ice mapping and discusses commonly used fusion strategies.
\hyperref[sec:dataset_preparation]{Section~III} describes the Arctic-SAR dataset preparation pipeline, including RCM scene matching, co-registration with Sentinel-1 observations, and dataset characterization. 
\hyperref[sec:methodology]{Section~IV} presents the methodology and our experimental configuration. 
\hyperref[sec:results]{Section~V} reports the experimental results and discusses the behavior of different fusion strategies, the contribution of RCM imagery, the limitations of the proposed benchmark, and future work. 
Finally, \hyperref[sec:conclusion]{Section~VI} concludes the article.

\section{Related Work}
\label{sec:related_work}

\subsection{Single-Sensor Learning for SAR-Based Sea Ice Mapping}
Deep learning has demonstrated effectiveness in automated sea ice mapping because it can learn spatial, textural, and semantic representations directly from SAR imagery \cite{Stokholm2022,WangLi2021}. 
Kortum et al. investigated robust multiseason sea ice classification from high-resolution TerraSAR-X X-band SAR imagery using a CNN-based approach, showing the feasibility of single-sensor learning across diverse seasonal conditions \cite{Kortum2022}. Khaleghian et al. evaluated several CNN architectures for both binary ice-water discrimination and multiclass sea ice classification using Sentinel-1 SAR imagery \cite{Khaleghian2021}. Additionally, U-Net-based architectures have been particularly influential because their encoder--decoder design combines high-level contextual representations with fine-scale spatial information through skip connections \cite{Ronneberger2015}. Huang et al. developed a dual-branch encoder U-Net for sea-ice type classification from SAR imagery in the Beaufort Sea, improving the extraction and fusion of multiscale features for distinguishing different ice classes \cite{Huang2024}. Ren et al. proposed a dual-attention U-Net for sea ice and open-water classification from dual-polarization SAR imagery, incorporating spatial and channel attention mechanisms and using incidence angle as an additional input \cite{Ren2021}.

Most existing early deep learning studies focused on single-sensor SAR data, leveraging its all-weather imaging capability and rich textural information. Sentinel-1 forms the basis of many prominent sea ice classification, segmentation, and benchmarking studies \cite{deLoe2025,Khaleghian2021,Stokholm2022,Chen2024}, due to its systematic polar acquisitions, dual-polarization observations, and openly accessible Copernicus data products provided by the European Space Agency \cite{ESA_Sentinel1}. However, single-sensor approaches are inherently influenced by sensor-specific patterns, which may limit model transferability across platforms.

\subsection{Cross-Sensor and Multimodal Fusion for SAR-Based Sea Ice Mapping}
To address the inherent limitations of single-sensor observations, previous studies have investigated multimodal and cross-sensor fusion strategies to integrate complementary information from different remote sensing sources. Han et al. applied two separate deep learning models to Sentinel-1 SAR and Sentinel-2 multispectral imagery and fused the extracted spatial and spectral features for sea ice classification \cite{Han2021}. Khachatrian et al. proposed an adaptive graph-Laplacian-based fusion scheme for combining Sentinel-1 SAR imagery with AMSR-2 passive microwave observations, demonstrating the complementary value of the high-resolution spatial information provided by SAR and the broader-scale radiometric information provided by passive microwave observations for sea ice classification \cite{Khachatrian2023}. These studies demonstrate the value of combining SAR imagery with complementary non-SAR sensing modalities for sea ice mapping. 

Compared with SAR-optical and SAR-passive microwave fusion, considerably fewer studies have investigated observations from multiple SAR missions. Johansson et al. proposed a semi-supervised method for consistent ice-water classification across imagery from ERS-1/2, Envisat ASAR, RADARSAT-2, and Sentinel-1 \cite{Johansson2020}. More recently, Gao et al. developed a keypoint-free feature-tracking method for Arctic sea ice motion retrieval and evaluated it on a multisource SAR test set comprising Sentinel-1, ALOS-2, C-SAR, Envisat ASAR, and RADARSAT-2 imagery \cite{Gao2025}. However, these studies primarily address cross-mission classification consistency, temporal analysis, or sea ice motion retrieval rather than paired cross-sensor fusion for dense sea ice mapping. 

In existing studies, cross-sensor and multimodal fusion methods are commonly categorized according to the stage at which information is integrated: input-level fusion, feature-level fusion, and decision-level fusion \cite{Xu2018, Ma2024, Pereira2024}.

\subsubsection{Input-Level Fusion}
Input-level fusion, also referred to as early fusion, combines observations from different sensors or modalities before feature extraction by concatenating the input channels and processing them using a shared encoder. This strategy enables the network to learn joint representations and potentially capture complementary relationships that may not be apparent from a single input source \cite{Pereira2024, deLoe2025}. For example, Huang et al. developed an end-to-end gated residual refinement network that fused high-resolution aerial imagery with LiDAR point clouds at the input stage, thereby combining complementary spectral and structural cues for building extraction \cite{Huang2019}.

\subsubsection{Feature-Level Fusion}
Feature-level fusion processes each sensor or modality through an independent encoder and integrates the resulting intermediate representations at one or more stages of the network. By preserving independent feature-extraction pathways, this strategy allows each encoder to learn sensor-specific characteristics before integration, which can reduce interference between heterogeneous inputs during early processing \cite{Wei2024MGFNet, Liu2024SoftFormer, Ma2026GLFFuse, deLoe2025}. Rogers et al. used parallel encoders to extract features independently from SAR and multispectral imagery acquired at different spatial resolutions, and subsequently fused the learned representations for sea ice detection \cite{Rogers2024}.

\subsubsection{Decision-Level Fusion}
Decision-level fusion processes each sensor or modality using an independent prediction pipeline and combines the resulting outputs to produce the final estimate. The individual predictions may be integrated through simple or weighted averaging, hard-decision voting, and probabilistic inference. Because each branch performs complete feature extraction and prediction before fusion, this strategy preserves sensor-specific processing and can reduce interference between heterogeneous inputs
\cite{Pereira2024, Xiong2023}. In sea ice applications, Wang et al. used a Bayesian framework to combine ice--water information derived from SAR imagery with AMSR-E daily sea ice concentration products, improving the spatial detail and accuracy of the resulting SIC estimates \cite{WangDecisionFusion}. However, maintaining multiple encoder--decoder pipelines increases computational cost, and the final result may remain sensitive to noisy or inaccurate predictions from individual branches.

\section{Arctic-SAR Dataset Preparation}
\label{sec:dataset_preparation}
In this study, co-registration and fusion refer to distinct processing
stages. Co-registration denotes the spatial alignment of Sentinel-1 and RCM
imagery onto a common pixel grid to form paired observations, whereas fusion
refers to the integration of information from the two sensors within the
learning framework.
This section provides an overview of the input data used in the Arctic-SAR dataset and describes the construction pipeline, including RCM scene retrieval, radiometric calibration, co-registration to the Sentinel-1 grid, and characterization of the dataset’s geographic, temporal, and radiometric properties.

\begin{figure*}[t]
    \centering
    \includegraphics[width=\textwidth]{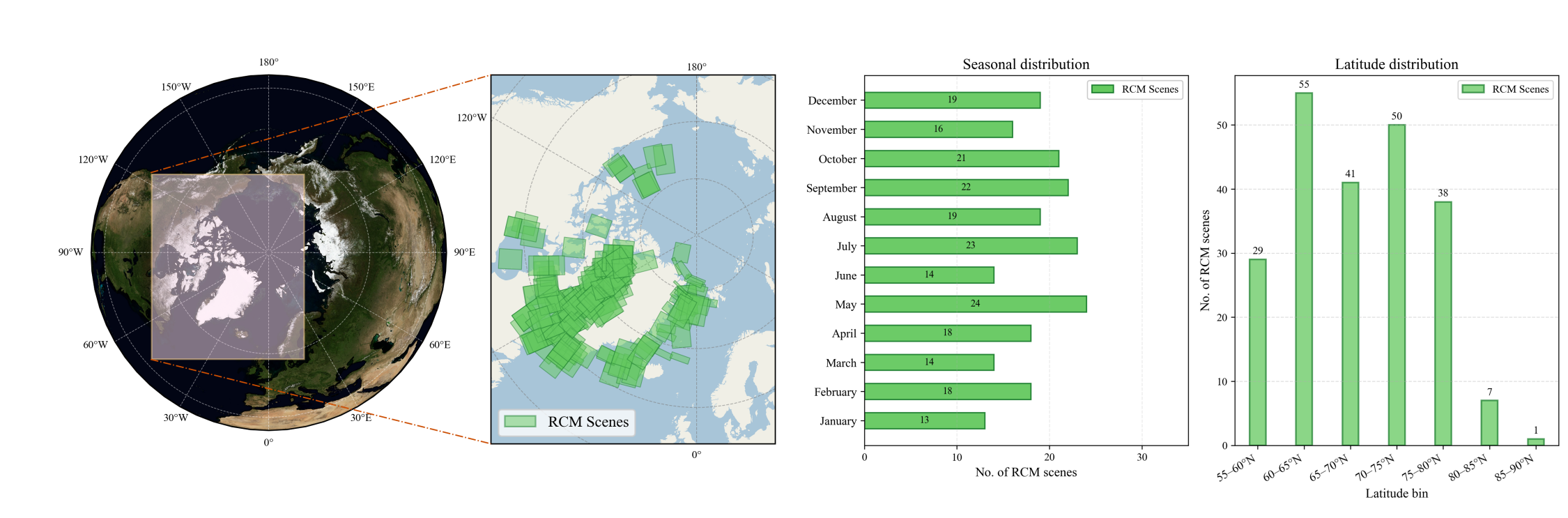}
    \caption{Geographical coverage and statistical distribution of the Arctic-SAR dataset. The figure shows the spatial distribution of RCM scenes over the Arctic region, together with the seasonal and latitudinal distributions.}
    \label{fig:dataset_distribution}
\end{figure*}

\subsection{AI4Arctic Data} 
The AI4Arctic dataset is a publicly available benchmark for SAR-based sea ice mapping in the Arctic~\cite{BuusHinkler2022}. It consists of 533 geo-referenced scenes acquired over the Canadian and Greenlandic Arctic regions between January 2018 and December 2021. Each scene includes Sentinel-1 Extra Wide (EW) swath imagery acquired in HH/HV dual polarization, ancillary data, AMSR2 passive microwave radiometer measurements, numerical weather prediction (NWP) parameters from the ERA5 reanalysis dataset, and ice chart annotations from the Canadian and Greenland Ice Services, which provide labels for sea ice concentration (SIC), stage of development (SOD), and floe size (FLOE). In this study, the Sentinel-1 SAR channels and ice-chart-derived labels from the raw AI4Arctic dataset are used as the baseline data source. Each scene includes geolocation and acquisition time information.
\begin{itemize}
    \item \textit{Sentinel-1 SAR (HH/HV):} The dataset includes dual-polarization Sentinel-1 backscatter imagery acquired in Extra Wide (EW) swath mode as Level-1 Ground Range Detected (GRD) products, with an approximate pixel spacing of 40~m. Thermal noise is corrected using the NERSC denoising algorithm~\cite{Korosov2022}, and backscatter values are represented as sigma-naught ($\sigma^0$) in decibels (dB).

    \vspace{0.3em}
    \item \textit{Ancillary Data:} Each data file contains auxiliary information, including latitude and longitude grids, scene acquisition time, incidence angle, and distance to the nearest coast. These auxiliary variables are aligned with the SAR channels and provided at the same pixel spacing.

    \item \textit{Ice Charts:} Expert-generated ice charts from the Canadian Ice Service (CIS) and the Danish Meteorological Institute (DMI) are temporally and geographically aligned with the corresponding SAR scenes and co-registered to the SAR grid. Ice charts follow the World Meteorological Organization (WMO) Egg Code format \cite{manice2005}, encoding three primary geophysical quantities within each polygon. Ice-chart polygons may contain multiple ice types and floe sizes. The associated partial concentrations are normalized by the total sea ice concentration to determine the dominant SOD and FLOE classes. Dominant classes are identified using a threshold of 65\%. When a dominant class is identified, the corresponding polygon is assigned a single SOD or FLOE label. Polygons without a dominant SOD or FLOE class are masked for the corresponding task \cite{Chen2024, Stokholm2024}. 

    \begin{itemize}
        \vspace{0.3em}
        \item[$\circ$] \textbf{Sea ice concentration (SIC):} Represents the fraction of ice-covered area within a region, ranging from 0\% for open water to 100\% for complete ice cover. SIC is stored as an integer from 0 to 10, corresponding to discrete 10\% intervals.
        \vspace{0.3em}
        \item[$\circ$] \textbf{Stage of development (SOD):} Describes the stage of sea ice development, which is related to ice thickness and strength. It is represented in six categories: Open Water (0), New Ice (1), Young Ice (2), Thin First-Year Ice (3), Thick First-Year Ice (4), and Old Ice (5).
        \vspace{0.3em}
        \item[$\circ$] \textbf{Floe size (FLOE):} Describes the spatial continuity and extent of sea ice coverage. It is represented in seven categories: Open Water (0), Cake Ice (1), Small Floe (2), Medium Floe (3), Big Floe (4), Vast Floe (5), and Bergs (6), where Bergs include icebergs and glacier ice.
    \end{itemize}
\end{itemize}

\subsection{RADARSAT Constellation Mission (RCM) Data}
The dataset includes 221 RCM dual-polarization SAR products. As shown in Fig.~\ref{fig:dataset_distribution}, the scene coverage is mainly concentrated over Baffin Bay and Davis Strait, with additional scenes distributed across the Canadian Arctic Archipelago and the Greenland Sea. The RCM scenes were selected to spatially and temporally overlap the corresponding AI4Arctic Sentinel-1 scenes. The selected RCM products are C-band Ground Range Detected (GRD) ScanSAR images acquired in the SC50MC beam mode, providing dual-polarization HH/HV backscatter at approximately 50~m pixel spacing with a wide-swath coverage of about 350~km. We then applied a preprocessing pipeline to convert the raw RCM imagery into model-ready inputs. This pipeline includes radiometric calibration of the HH and HV channels to the backscatter coefficient of $\sigma^0$ and geometric alignment with the corresponding Sentinel-1 scene through co-registration. As a result, the extended AI4Arctic dataset includes two additional aligned input channels for each paired scene. Using this extension, this study aims to quantify the contribution of the RCM channels to SAR-based sea ice mapping.

\vspace{0.3em}
\subsubsection{Data Retrieval and Scene Selection}
RCM SAR scenes were retrieved programmatically using the Earth Observation Data Management System (EODMS) API, the official data access portal for Canadian government satellite assets~\cite{EODMS}. The EODMS API was queried using the Sentinel-1 scene footprint to constrain the spatial search and the buffered acquisition times to constrain the temporal search. 
The returned candidate metadata were filtered to include only RCM products with dual-polarization HH/HV channels, ScanSAR Wide beam mode with a 350~km swath, GRD product format, and an ice-optimized radiometric look-up table (LUT). 

\paragraph{Spatial Overlap Assessment}
The spatial overlap between each RCM candidate and the Sentinel-1 scene was then calculated using polygon intersection. The Sentinel-1 footprint was derived from the previously extracted shapefiles, and the RCM footprint was generated from geolocation tie-point grids in the product XML metadata. The candidate with the greatest overlap was selected for subsequent processing.

\paragraph{Temporal Matching and Acquisition-Time Differences}
For each Sentinel-1 scene, acquisition timestamps were extracted from the scene filename, and a symmetric $\pm 5$~hour search interval, corresponding to a 10~hour temporal search window, was used to identify candidate RCM acquisitions around each Sentinel-1 acquisition. Across the 221 selected Sentinel-1--RCM pairs, the mean absolute acquisition-time difference was 18.9~min, the median was 16.75~min, and 95\% of the pairs had a time difference below 47.83~min. The maximum observed time difference was 3.23~hours (194.05~min). This interval was chosen because wider temporal windows produced RCM--Sentinel-1 pairs with greater spatial differences in observed sea ice conditions. The search interval was therefore narrowed to improve temporal consistency between paired acquisitions.

\paragraph{Potential Sea-Ice Displacement}
To estimate the potential spatial mismatch caused by the acquisition-time gap between the paired Sentinel-1 and RCM acquisitions, we used the Copernicus Marine Service (CMEMS) Global Ocean--High Resolution SAR Sea-Ice Drift Time Series product~\cite{CMEMSSeaIceDriftData}. The CMEMS 24-hour composite provides the mean horizontal sea-ice displacement components, $dX_{\mathrm{mean}}$ and $dY_{\mathrm{mean}}$, on a 10-km polar stereographic grid, with the displacement components normalized to a 24-hour interval~\cite{CMEMSSeaIceDrift}. Following the displacement and speed definitions~\cite{QiuLi2026}, the local CMEMS displacement magnitude was calculated as
\begin{equation}
D_{\mathrm{CMEMS}} =
\sqrt{
dX_{\mathrm{mean}}^{2}
+
dY_{\mathrm{mean}}^{2}
},
\end{equation}
and the corresponding mean sea-ice drift speed was obtained as
\begin{equation}
S_{\mathrm{ice}} =
\frac{D_{\mathrm{CMEMS}}}{T_{\mathrm{CMEMS}}},
\end{equation}
where $T_{\mathrm{CMEMS}}=24$~hours denotes the temporal interval represented by the CMEMS displacement fields. Assuming this mean drift speed is representative of the shorter interval between the SAR acquisitions, the potential displacement over their acquisition-time difference is estimated as:
\begin{equation}
D_{\mathrm{gap}} =
S_{\mathrm{ice}}
\,\Delta t_{\mathrm{S1-RCM}}.
\end{equation}

Only valid CMEMS grid cells located within the common spatial footprint of the corresponding Sentinel-1 and RCM scenes were considered, and the median local displacement was used as the representative displacement for each scene pair. Of the 221 Sentinel-1--RCM scene pairs, 80 had valid CMEMS sea-ice drift coverage over their common spatial footprint and were included in the displacement analysis. Across these 80 scene pairs, the estimated scene-level displacement had a mean of $142.4$~m, while 95\% of the scene-level displacement estimates were below $463.3$~m. The largest representative scene-level median displacement was $2.32$~km, while the largest local displacement observed across all valid CMEMS grid cells was $2.38$~km. These results indicate relatively limited estimated displacement for most evaluated pairs, although substantially larger local motion occurred in a small number of cases.

\vspace{0.3em}
\subsubsection{Preprocessing and Radiometric Calibration}
After retrieval, the raw HH and HV GeoTIFF bands were converted from 16-bit integer values to float32. Zero-valued pixels were then replaced with NaN to correctly handle no-data regions during processing. Radiometric calibration from digital numbers (DNs) to $\sigma^0$ backscatter in dB was performed following the RCM Image Product Format Definition \cite{RCMProductFormat}, using the polarization-specific lookup tables (LUTs) provided in the RCM product metadata (\texttt{lutSigma-HH.xml} and \texttt{lutSigma-HV.xml}). Each LUT provides a one-dimensional array of gain values along the range direction with a fixed sampling interval, and a constant offset $B$. Since the gain is not specified at every range pixel, the discrete gains are linearly interpolated to cover the full range extent of the image. The resulting one-dimensional gain profile is then applied uniformly to all azimuth lines to form a two-dimensional gain field. Following
\cite{RCMProductFormat}, $\sigma^0$  in linear scale was computed as:
\begin{equation}
\sigma^{0}_{\mathrm{lin}}
=
\frac{\mathrm{DN}^{2}+B}{\mathrm{gain}}.
\end{equation} For the RCM
products used in this study, the LUT offset was zero.
The calibrated values were then converted to decibels as: 
\begin{equation}
\sigma^0_{\mathrm{dB}} = 10 \log_{10}\left(\sigma^0_{\mathrm{lin}}\right).
\end{equation}
The calibrated HH and HV images were saved as float32 GeoTIFFs. The calibrated RCM backscatter distributions were then compared with the corresponding Sentinel-1 channels using the normalized backscatter intensity distributions, as shown in Fig.~\ref{fig:backscatter_distribution}, to assess their cross-sensor radiometric consistency.

\vspace{0.3em}
\subsubsection{Co-registration with Sentinel-1}
Co-registration of the calibrated RCM imagery onto the Sentinel-1 pixel grid is performed using a two-stage geolocation-based geometrical registration method. Following the reference--secondary framework used in
geometrical SAR image co-registration \cite{Li2015, Sansosti2006}, Sentinel-1 is
treated as the reference image and RCM as the resampled image. The procedure uses the sparse geolocation information (tie-point arrays) provided by both sensors to determine the corresponding RCM image coordinates for each Sentinel-1 pixel and subsequently resample the RCM imagery onto the Sentinel-1 grid. The overall mapping can be expressed as:

\begin{equation}
(r,c)_{\mathrm{S1}}
\rightarrow
(\phi,\lambda)
\rightarrow
(u,v)_{\mathrm{RCM}}
\rightarrow
I_{\mathrm{RCM}}^{\mathrm{co-reg}}(r,c),
\end{equation} where $(r,c)_{\mathrm{S1}}$ denotes a Sentinel-1 pixel location,
$(\phi,\lambda)$ its geographic latitude and longitude,
$(u,v)_{\mathrm{RCM}}$ the corresponding fractional RCM row and column coordinates,
and $I_{\mathrm{RCM}}^{\mathrm{co-reg}}(r,c)$ the RCM backscatter value after bilinear resampling onto the Sentinel-1 reference grid.

\vspace{0.3em}
\textit{Stage 1 - Dense Sentinel-1 geolocation:}
The sparse Sentinel-1 geolocation grid, consisting of tie-point line and sample coordinates mapped to geographic latitude and longitude, is interpolated across the full Sentinel-1 scene using cubic scattered interpolation. This produces dense latitude and longitude arrays at the full Sentinel-1 image resolution, defining the geographic location of each Sentinel-1 pixel.

\vspace{0.3em}
\textit{Stage 2 - Inverse mapping to RCM coordinates:}
The RCM tie-point grid provides the forward mapping from RCM image space, represented by line and sample coordinates, to geographic latitude and longitude. These tie points are used as scattered control points to interpolate the corresponding RCM row and column indices at each Sentinel-1 pixel location. This produces two coordinate maps, one for the RCM row index and one for the RCM column index, defining the fractional RCM image position corresponding to each Sentinel-1 pixel. The calibrated RCM HH and HV images are then resampled onto the Sentinel-1 grid by bilinear interpolation, placing both sensors on a common spatial grid. Nearest-neighbor extrapolation is then applied for Sentinel-1 pixels that fall outside the RCM scene extent, producing co-registered RCM arrays with the same dimensions as the Sentinel-1 images. Pixels outside the common valid Sentinel-1/RCM footprint were excluded from subsequent analysis, training, and evaluation. The co-registered RCM outputs are written as float32 GeoTIFF files with ground control points (GCPs) derived from the Sentinel-1 sparse geolocation grid attached directly to the file, using the WGS-84 geographic coordinate reference system (EPSG:4326). This ensures that the co-registered RCM channels inherit the spatial reference framework of the Sentinel-1 scene.

\vspace{0.5em}
\noindent\textit{Co-registration accuracy assessment:} To quantitatively assess the residual co-registration accuracy between the Sentinel-1 and RCM imagery, a checkpoint-based evaluation was performed using established SAR image registration accuracy metrics \cite{Tang2024}. The co-registration accuracy assessment was performed on 161 of the 221 Sentinel-1--RCM scene pairs. For each evaluated pair, candidate checkpoints were randomly sampled across the image, automatically matched, and visually verified, after which 20 reliable and spatially distributed corresponding checkpoints were selected for accuracy assessment. The remaining 60 scene pairs were excluded because a sufficient number of reliable corresponding checkpoints could not be identified and verified. The residual error for the $i$th checkpoint was calculated as \cite{Tang2024}:

\begin{equation}
e_i =
\sqrt{
(x_i-x_i')^2 +
(y_i-y_i')^2
},
\label{eq:coreg_error}
\end{equation}

\noindent
where $(x_i,y_i)$ and $(x_i',y_i')$ denote the corresponding checkpoint
coordinates in the Sentinel-1 and co-registered RCM images, respectively.
Following \cite{Tang2024}, the root-mean-square error (RMSE) and standard deviation (STD) of the checkpoint errors were computed as:

\begin{equation}
\mathrm{RMSE} =
\sqrt{
\frac{1}{m}
\sum_{i=1}^{m}
\left[
(x_i-x_i')^2 +
(y_i-y_i')^2
\right]
},
\label{eq:coreg_rmse}
\end{equation}

\begin{equation}
\mathrm{STD} =
\sqrt{
\frac{1}{m}
\sum_{i=1}^{m}
\left(e_i-\bar{e}\right)^2
},
\label{eq:coreg_std}
\end{equation}
where $m=20$ is the number of checkpoints per evaluated image pair and $\bar{e}$ is the mean checkpoint error. In total, 3,220 checkpoints were included in the assessment. The mean scene-level RMSE was $6.60$ pixels, corresponding to approximately $264$~m on the Sentinel-1 grid with an approximate pixel spacing of $40$~m. The mean scene-level STD was $3.34$ pixels. These results quantify the residual cross-sensor
correspondence uncertainty after co-registration, part of which may also reflect temporal changes in the sea ice surface between the two acquisitions.

\vspace{0.3em}
\subsubsection{Dataset Integration}

\begin{figure*}[t]
    \centering
\includegraphics[width=\textwidth]{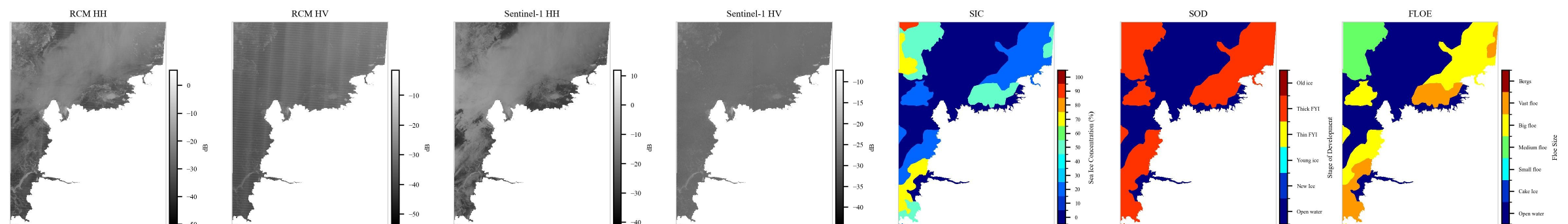}
    \caption{Example paired scene from Arctic-SAR showing co-registered RCM HH/HV and Sentinel-1 HH/HV channels and the corresponding ice-chart-derived labels for sea ice concentration (SIC), stage of development (SOD), and floe size (FLOE). RADARSAT Constellation Mission Imagery \textcopyright\ Government of Canada (2019–2021). RADARSAT is an official mark of the Canadian Space Agency.}
    \label{fig:sar_visualization}
\end{figure*}

The resulting Arctic-SAR dataset adds two aligned input channels (RCM HH/HV) to each of the 221 Sentinel-1 scenes. The dataset is organized in NetCDF4 format~\cite{Unidata2020NetCDF}, inheriting the AI4Arctic file structure and variable naming conventions. The two co-registered RCM channels, \texttt{sar\_RCM\_HH\_cor\_cal} and \texttt{sar\_RCM\_HV\_cor\_cal}, are appended to the corresponding AI4Arctic NetCDF files as new variables alongside the Sentinel-1 channels and ice-chart labels, as shown in Fig.~\ref{fig:sar_visualization}. The original native-grid RCM arrays, \texttt{orig\_sar\_RCM\_HH} and \texttt{orig\_sar\_RCM\_HV}, are also preserved for reference. Additionally, RCM incidence angle \texttt{sar\_grid\_incidenceangle\_rcm}, and spatiotemporal metadata, including latitude, \texttt{sar\_grid\_latitude\_rcm}, longitude, \texttt{sar\_grid\_longitude\_rcm}, and sample/line coordinates, \texttt{sar\_grid\_sample\_rcm} and \texttt{sar\_grid\_line\_rcm}, were added as new variables to each file. The 221 paired scenes in Arctic-SAR were partitioned for five-fold cross-validation at the scene level, as described in Section~\ref{sec:experimental_configurations}. It should be noted that, unlike the Sentinel-1 channels in AI4Arctic, which use the NERSC thermal-noise correction algorithm, the RCM channels in Arctic-SAR were not processed with an equivalent noise-floor subtraction, as the Sentinel-1-specific procedure cannot be directly transferred to RCM and an RCM-specific correction requires separate development and validation, which is left for future work.

\subsection{Dataset Statistics}

\subsubsection{Label Distribution}
\begin{figure}[!t]
    \centering
\includegraphics[width=\columnwidth]{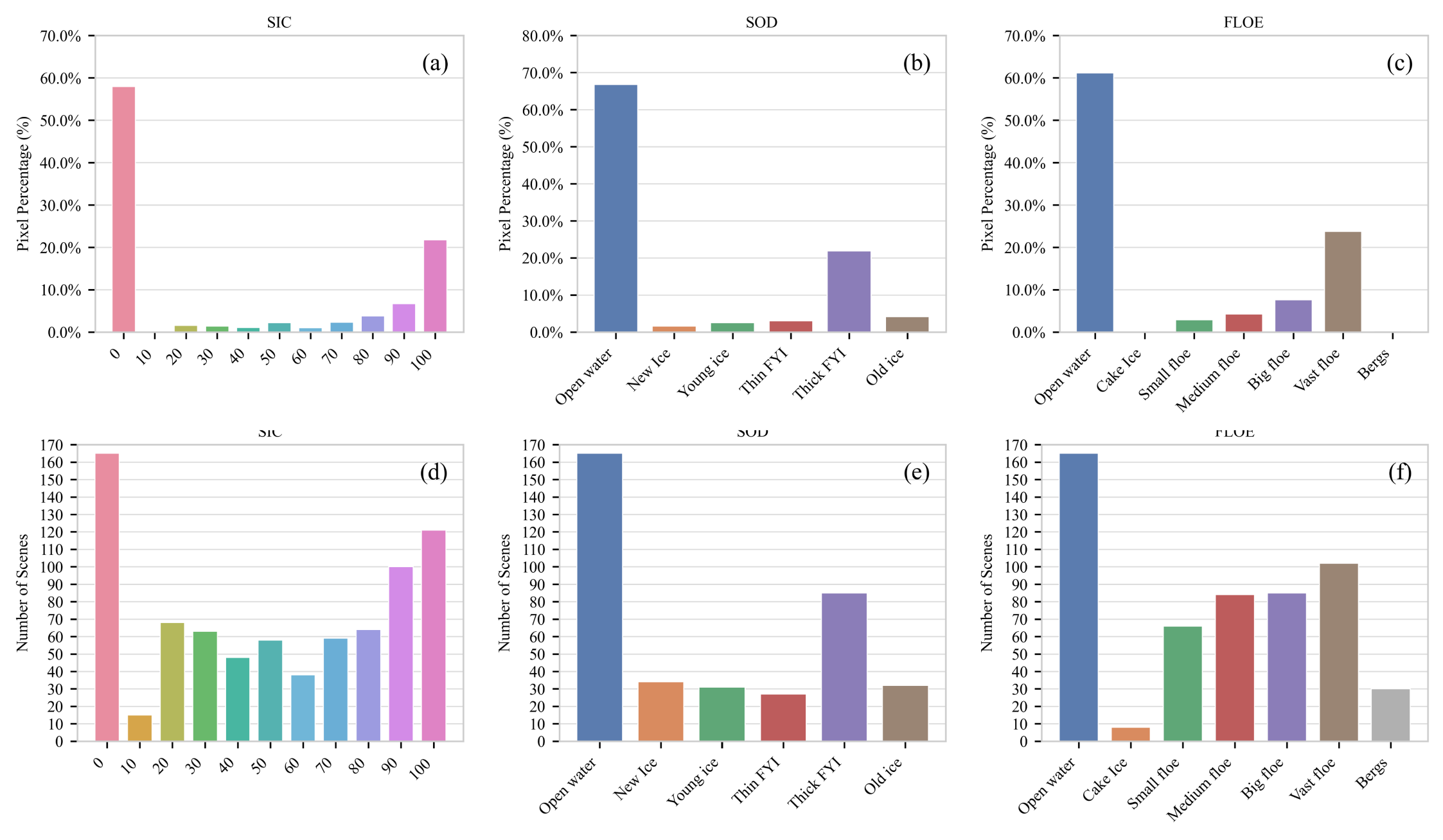}
    \caption{Label distribution statistics for the Arctic-SAR dataset. Panels (a)--(c) show the pixel-level percentage distributions of the SIC, SOD, and FLOE classes, respectively. Panels (d)--(f) show the corresponding number of scenes in which each ice-chart-derived class is represented.}
    \label{fig:label_distribution}
\end{figure}

Fig.~\ref{fig:label_distribution} summarizes the pixel-level label distributions and scene-level class coverage across the 221 paired scenes in Arctic-SAR. Panels (a)--(c) show the pixel-level percentage of each SIC, SOD, and FLOE class pooled across all scenes, while panels (d)--(f) show the number of scenes containing at least one labeled pixel for each class. Overall, the dataset exhibits clear class imbalance across all three tasks. For SIC, the distribution is strongly concentrated at the 0\% and 100\% bins, indicating that open water and full ice cover dominate the labeled pixels, while intermediate concentration values are less frequent. However, the scene-count distribution shows that these intermediate SIC classes still appear across many scenes, suggesting broad spatial coverage despite limited pixel representation. For SOD and FLOE, Open Water is again the dominant pixel-level class. Among ice classes, Thick First-Year Ice is the most frequent SOD category, while Vast Floe dominates the FLOE labels. Less frequent categories, including New Ice, Young Ice, Thin First-Year Ice, Old Ice, Cake Ice, and Bergs, occupy relatively small pixel fractions but are present in multiple scenes. This indicates that the dataset includes diverse sea-ice conditions, although their spatial extent is highly uneven. These imbalanced but physically meaningful label distributions should be considered when training and evaluating multi-task sea ice segmentation models, particularly for minority SOD and FLOE classes.

\begin{figure*}[t]
    \centering
    \includegraphics[width=\textwidth]{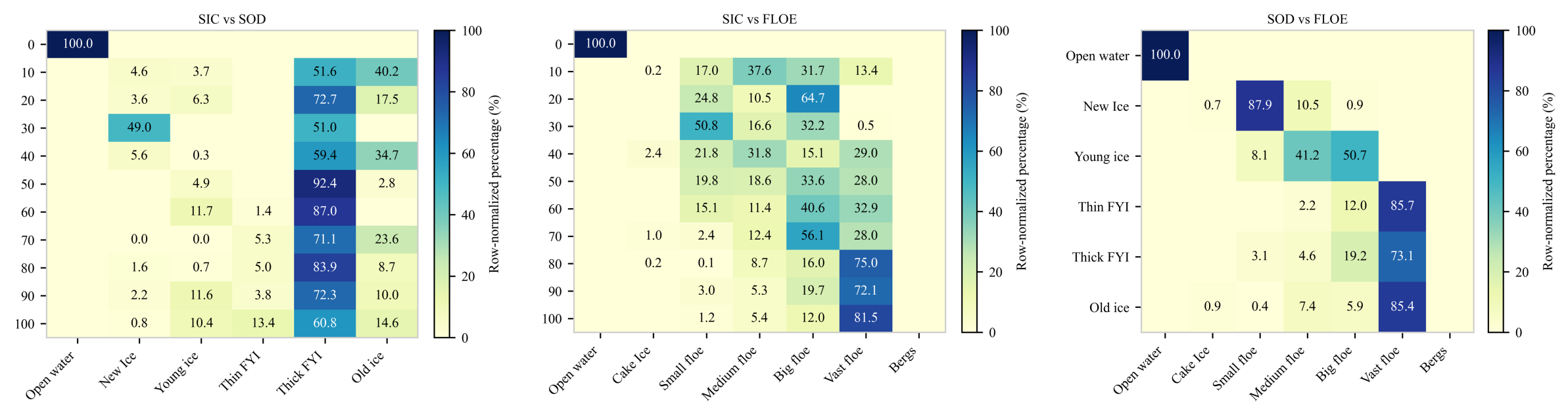}
    \caption{Label correlation maps for the Arctic-SAR dataset, showing the row-normalized percentage relationships between SIC--SOD, SIC--FLOE, and SOD--FLOE.}
    \label{fig:label_correlation}
\end{figure*}

Fig.~\ref{fig:label_correlation} further analyzes row-normalized label correlation maps between the three label tasks. Each row represents a class from one task, and the values indicate how pixels in that class are distributed across classes of another task. The maps show strong physical consistency among the ice-chart-derived labels. The SIC--SOD correlation map shows that 0\% SIC is exclusively associated with open water, while non-zero SIC classes are mainly linked to ice-development categories, particularly Thick First-Year Ice. Old Ice also appears across several SIC bins, whereas New Ice and Young Ice are more prominent at lower or intermediate concentrations. The SIC--FLOE map shows that 0\% SIC maps entirely to open water, while increasing SIC values are generally associated with larger floe categories. Lower and intermediate SIC bins contain a mixture of Small, Medium, and Big Floe, whereas higher SIC bins are increasingly dominated by Vast Floe. The SOD--FLOE map further shows a physically meaningful relationship between ice type and floe size: New Ice is mainly associated with Small Floe, Young Ice with Medium and Big Floe, and Thin First-Year Ice, Thick First-Year Ice, and Old Ice predominantly with Vast Floe. Overall, these correlations indicate that SIC, SOD, and FLOE describe related but complementary aspects of sea ice conditions. This relationship provides useful context for multi-task learning, where shared physical structure across labels can support joint learning and help interpret class-specific errors.

\subsubsection{Dataset Distribution}
Fig.~\ref{fig:dataset_distribution} summarizes the geographic, seasonal, and latitudinal distribution of the 221 RCM scenes in Arctic-SAR. The scene footprints show broad coverage across the Arctic sea ice zone, with most scenes located around the Canadian Arctic Archipelago, Baffin Bay, Hudson Bay, Greenland, and adjacent Arctic waters. The monthly distribution indicates near year-round coverage, with scenes available in all months. The latitude distribution shows that most scenes fall between approximately 60$^\circ$N and 80$^\circ$N, with fewer scenes at higher latitudes.

\subsubsection{Cross-Sensor Radiometric Characterization}
\begin{figure}[t]
    \centering
    \includegraphics[width=\columnwidth]{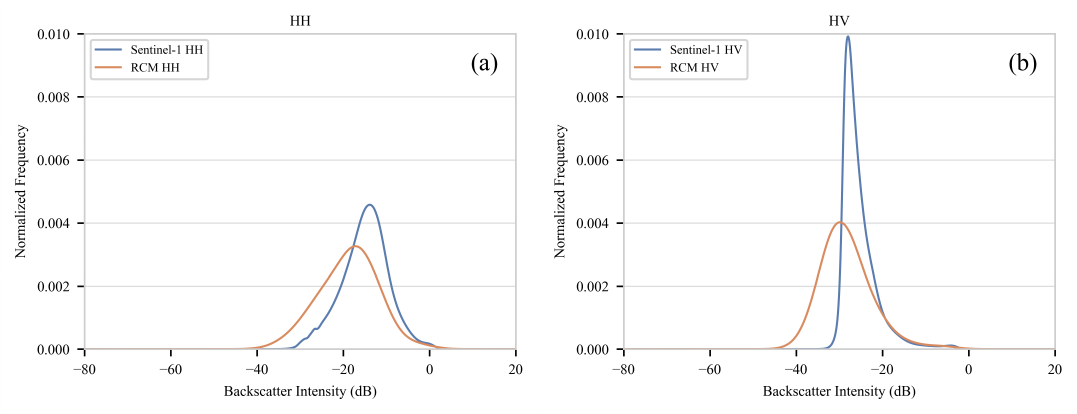}
    \caption{Backscatter intensity distributions of co-registered RCM and Sentinel-1 SAR channels. Panel (a) compares the HH backscatter distributions, while panel (b) compares the HV backscatter distributions. The distributions show that Sentinel-1 and RCM have broadly similar dynamic ranges, but RCM exhibits a broader distribution, particularly in the HV channel, indicating sensor-specific radiometric differences.}
    \label{fig:backscatter_distribution}
\end{figure}

\begin{figure}[t]
    \centering
    \includegraphics[width=\columnwidth]{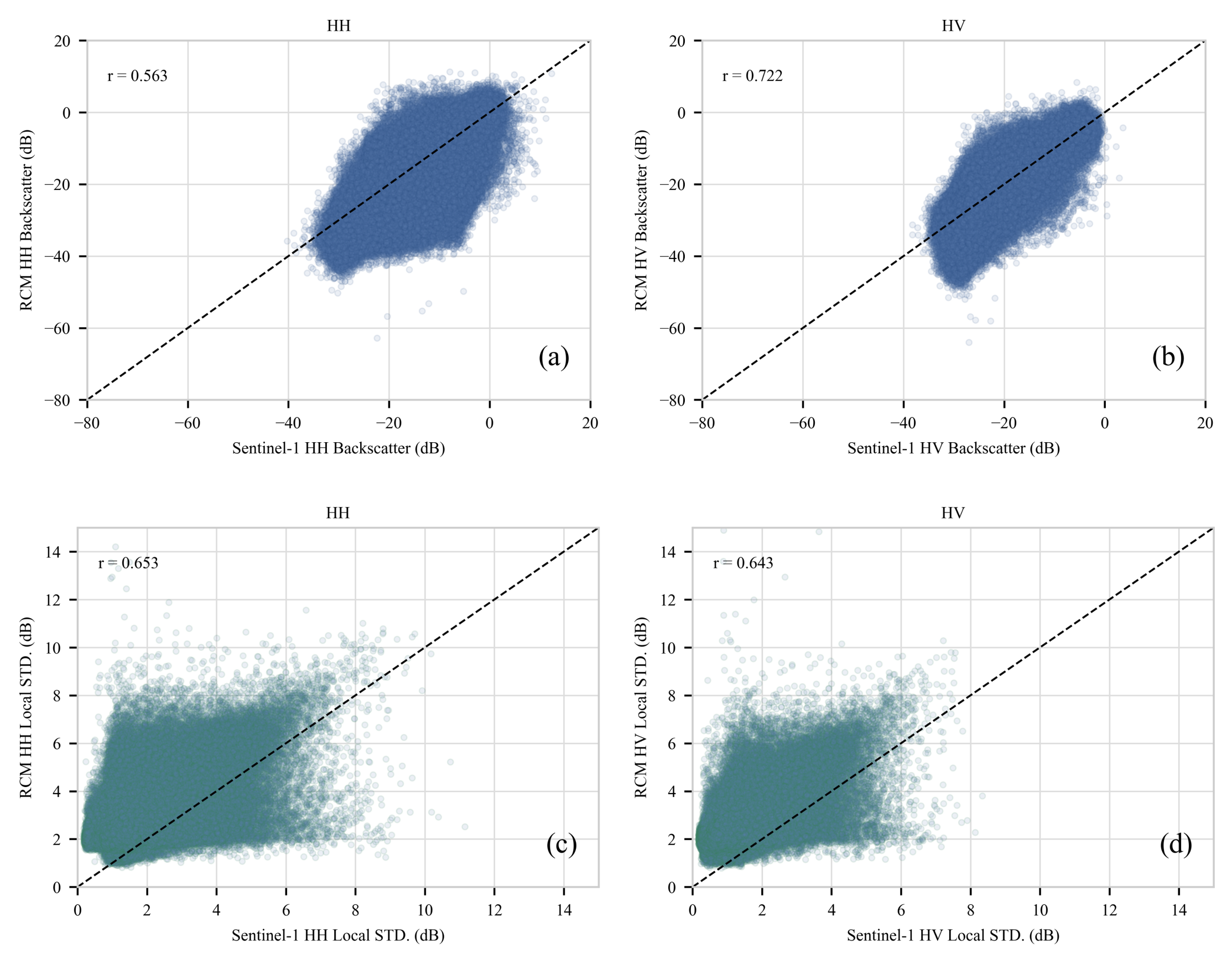}
    \caption{Cross-sensor comparison of Sentinel-1 and RCM over all common valid regions for HH and HV polarization channels. Panels (a) and (b) show the pixel-level backscatter intensity relationships, while panels (c) and (d) show the corresponding local backscatter variability, measured using the standard deviation within an $11\times11$ window. Each point represents a paired Sentinel-1--RCM value after co-registration. Pearson's correlation coefficient ($r$) is reported in each panel, and the dashed line indicates the one-to-one relationship.}
    \label{fig:cross_sensor_scatter}
\end{figure}

A critical aspect of cross-sensor dataset design is characterizing the degree of radiometric consistency between the contributing sensors, as corresponding observations from different sensors need to be radiometrically comparable for reliable fusion. Radiometric differences arising from sensor characteristics, acquisition conditions, and viewing geometry can reduce this comparability, while radiometric intercalibration is used to improve cross-sensor consistency and support reliable multi-sensor fusion \cite{Jiang2022, Samadzadegan2025}. This is also important for deep learning, as cross-sensor SAR variations can alter scattering intensity and feature distributions, potentially reducing model generalization across sensors \cite{Zhang2026CrossSensor}. Sentinel-1 and RCM both operate in the C-band and provide dual-polarization HH/HV observations, but differ in their noise floors, calibration chains, incidence-angle distributions, and temporal sampling \cite{ESA_Sentinel1, RCMProductFormat, Korosov2022}. Figs.~\ref{fig:backscatter_distribution} and \ref{fig:cross_sensor_scatter} present a comprehensive cross-sensor radiometric analysis of the paired HH and HV channels.

\textit{Backscatter Intensity Distributions:} 
Fig.~\ref{fig:backscatter_distribution} compares the backscatter intensity distributions of Sentinel-1 (blue) and RCM (orange) for the HH and HV polarization channels over their common valid regions across all paired scenes. For both polarizations, the two sensors show broadly similar intensity ranges, approximately from $-40$ to 0~dB. However, the RCM distributions are slightly broader and extend toward lower backscatter values than Sentinel-1. This difference is more evident in the HV channel, where Sentinel-1 exhibits a sharper peak, while RCM shows a wider distribution with lower-intensity values. These differences indicate sensor-specific radiometric characteristics, potentially arising from differences in acquisition geometry, particularly incidence angle, noise behavior, calibration procedures, and the absence of an equivalent thermal-noise correction step for RCM. Overall, the distributions show that Sentinel-1 and RCM measure similar C-band backscatter ranges, but with clear sensor-dependent radiometric differences, especially in the HV channel. Thus, operating at the same frequency band does not necessarily imply redundant observations, as the two sensors can exhibit different radiometric responses \cite{Mouche2015Global}. These differences also indicate that future RCM-specific thermal-noise correction and sensor-aware fusion strategies may be necessary to better exploit the complementary information from both SAR sensors for sea ice mapping.

\vspace{0.3em}
\textit{Cross-Sensor Pixel-Level Correlation:}
Fig.~\ref{fig:cross_sensor_scatter} shows the pixel-level backscatter intensity and local backscatter variability relationships between co-registered Sentinel-1 and RCM observations for HH and HV polarizations. For the pixel-level backscatter in panels (a) and (b), both channels exhibit a positive correlation, indicating that brighter Sentinel-1 pixels generally correspond to brighter RCM pixels and confirming the physical consistency of the co-registered cross-sensor measurements. The HV channel shows a stronger correlation ($r=0.722$) than HH ($r=0.563$), suggesting that cross-polarized backscatter preserves more consistent relative contrast between the two sensors, despite being more sensitive to noise. However, the scatter around the one-to-one line indicates that the two sensors are not radiometrically identical. These differences may arise from non-simultaneous acquisitions, as sea ice movement between acquisitions can reduce pixel-wise correspondence, and may also result from incidence-angle differences and sensor-specific noise characteristics.

The local backscatter variability analysis in panels (c) and (d) is based on the local standard deviation computed within an $11\times11$ window, which measures the amount of backscatter variation within the surrounding neighborhood of each pixel. Low local standard deviation values correspond to relatively homogeneous local backscatter, whereas higher values reflect stronger local variability and spatial heterogeneity, including the presence of texture and edges \cite{Aiazzi2005Heterogeneity}. For both HH and HV, the local standard deviation measures are positively correlated ($r=0.653$ and $r=0.643$, respectively), indicating that the two sensors capture similar patterns of local spatial variability. However, most points lie above the one-to-one line, showing that RCM generally exhibits higher local standard deviation than Sentinel-1 in both polarizations. The broad scatter further shows that the magnitude of local variability differs between the two sensors. Overall, the results suggest similar patterns of local backscatter variability across the two sensors.

Although the two sensors exhibit moderate to strong correlations, these relationships suggest substantial shared C-band information rather than identical observations. The systematic offset from the one-to-one line and the remaining scatter indicate sensor-specific radiometric and spatial differences. However, these differences do not necessarily represent complementary information that improves sea-ice prediction.

In summary, Arctic-SAR provides a physically meaningful, well-characterized cross-sensor dataset that pairs Sentinel-1 and RCM SAR observations over a diverse range of Arctic sea ice conditions. The cross-sensor radiometric analysis confirms moderate-to-strong backscatter correlations between sensors while revealing systematic differences, particularly in the HV channel. These differences motivate the cross-sensor fusion experiments.

\section{Methodology}
\label{sec:methodology}
This section presents the proposed benchmark framework for systematically evaluating single-sensor and cross-sensor SAR learning strategies for automated Arctic sea ice mapping. Using \textit{Arctic-SAR} as described in \hyperref[sec:dataset_preparation]{Section~III}, which consists of 221 co-registered RCM and Sentinel-1  scenes, the benchmark is designed to assess the contribution of RCM observations and determine how the effectiveness of cross-sensor integration varies across fusion stages for the prediction of three operational sea ice parameters: sea ice concentration (SIC), stage of development (SOD), and floe size (FLOE). Five experimental configurations are evaluated: Sentinel-1-only learning, RCM-only learning, input-level fusion, feature-level fusion, and decision-level fusion. All models are formulated as multi-task dense prediction networks and generate one output map for each sea ice parameter.

\begin{figure*}[!t]
\centering
\includegraphics[width=\textwidth]{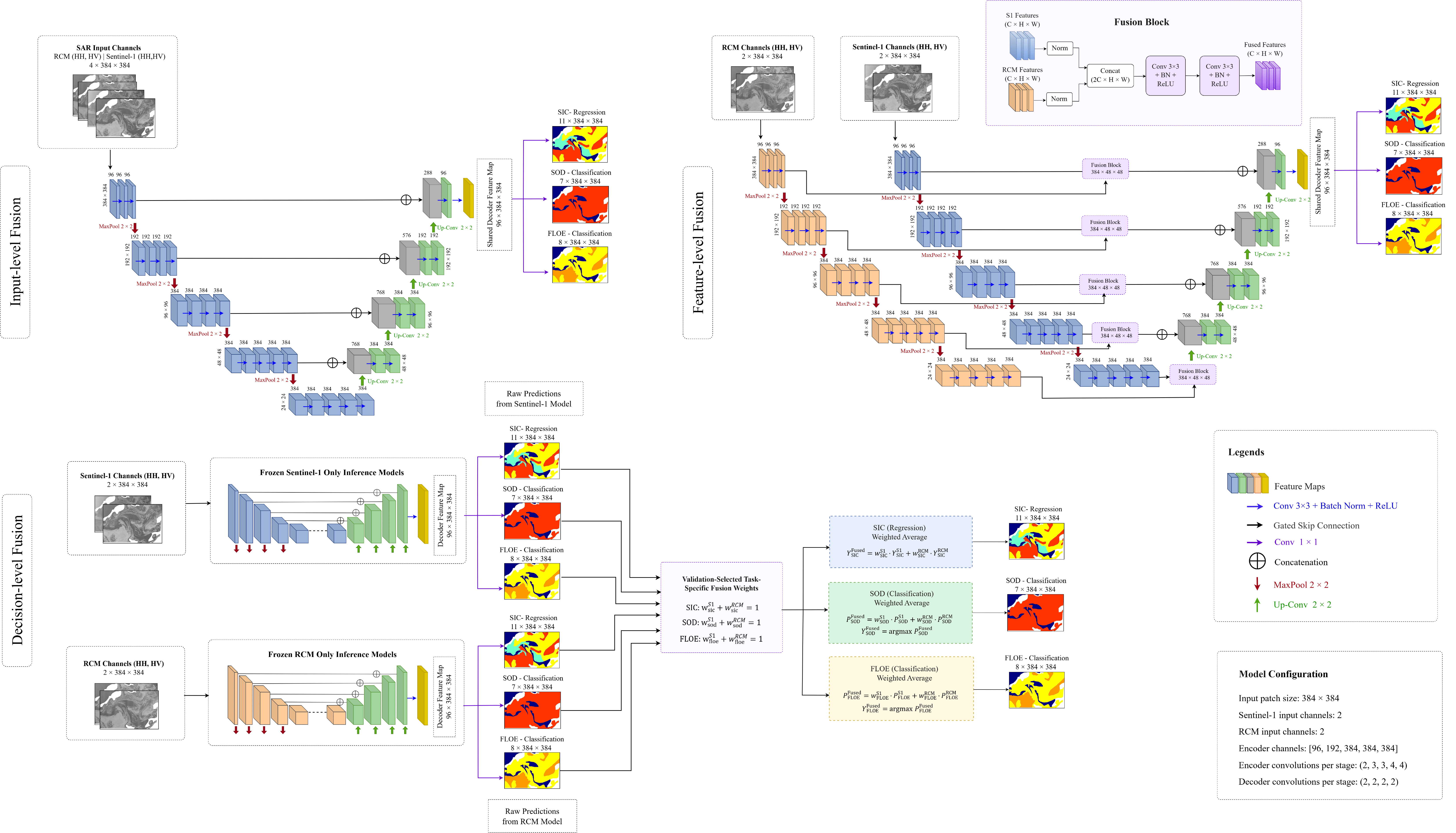}
\caption{
Overview of the three cross-sensor fusion strategies for SAR-based sea ice mapping, including input-level fusion, feature-level fusion, and decision-level fusion. The input-level fusion model concatenates Sentinel-1 and RCM channels before feature extraction and uses a single encoder--decoder network. The feature-level fusion model employs two sensor-specific encoders and combines Sentinel-1 and RCM representations through multi-scale fusion blocks before decoding. The decision-level fusion strategy uses frozen Sentinel-1-only and RCM-only inference models and combines their task-specific predictions using validation-selected fusion weights.
All configurations produce multi-task outputs for sea ice concentration (SIC), stage of development (SOD), and floe size (FLOE).
RADARSAT Constellation Mission Imagery
\textcopyright\ Government of Canada (2019–2021).
RADARSAT is an official mark of the Canadian Space Agency.}
\label{fig:model_architectures}
\end{figure*}

\subsection{Problem Formulation}
To formally define these benchmark configurations, we first introduce the notation for the input modalities and networks. We use $\mathbf{X}_{\mathrm{S1}} \in \mathbb{R}^{2 \times H \times W}$ and
$\mathbf{X}_{\mathrm{RCM}} \in \mathbb{R}^{2 \times H \times W}$ to denote the Sentinel-1 and RCM SAR HH/HV polarization input channels, respectively. For each input scene, the model predicts three dense output maps:
\begin{equation}
\hat{\mathcal{Y}}
=
\left\{
\hat{\mathbf{Y}}_{\mathrm{SIC}},
\hat{\mathbf{Y}}_{\mathrm{SOD}},
\hat{\mathbf{Y}}_{\mathrm{FLOE}}
\right\},
\end{equation} where $\hat{\mathbf{Y}}_{\mathrm{SIC}}$ is treated as a regression output, while
$\hat{\mathbf{Y}}_{\mathrm{SOD}}$ and $\hat{\mathbf{Y}}_{\mathrm{FLOE}}$
are treated as semantic classification outputs. The encoder and decoder components of the multi-task U-Net are represented by
$f^{e}(\cdot)$ and $f^{d}(\cdot)$, respectively. The operator $\Vert$
denotes channel-wise concatenation. In the feature-level fusion model,
$i \in \{1,\ldots,L\}$ indexes spatial scales of the encoder, and
$\psi_i(\cdot)$ denotes the fusion block that integrates the Sentinel-1
and RCM feature maps at the corresponding spatial scale $i$. Using this
notation, the five benchmark configurations are formally defined as follows:

\begin{enumerate}

\item \textbf{Sentinel-1-only learning:}
\begin{equation}
\hat{\mathcal{Y}}_{\mathrm{S1}}
=
f^{d}_{\mathrm{S1}}
\left(
f^{e}_{\mathrm{S1}}
\left(
\mathbf{X}_{\mathrm{S1}}
\right)
\right).
\end{equation}

\item \textbf{RCM-only learning:}
\begin{equation}
\hat{\mathcal{Y}}_{\mathrm{RCM}}
=
f^{d}_{\mathrm{RCM}}
\left(
f^{e}_{\mathrm{RCM}}
\left(
\mathbf{X}_{\mathrm{RCM}}
\right)
\right).
\end{equation}

\item \textbf{Input-level fusion:}
\begin{equation}
\hat{\mathcal{Y}}_{\mathrm{I}}
=
f^{d}_{\mathrm{I}}
\left(
f^{e}_{\mathrm{I}}
\left(
\mathbf{X}_{\mathrm{S1}}
\Vert
\mathbf{X}_{\mathrm{RCM}}
\right)
\right).
\end{equation}

\item \textbf{Feature-level fusion:}
\begin{equation}
\hat{\mathcal{Y}}_{\mathrm{F}}
=
f^{d}_{\mathrm{F}}
\left(
\left\{
\psi_i
\left(
f^{e}_{\mathrm{S1},i}
\left(
\mathbf{X}_{\mathrm{S1}}
\right)
\Vert
f^{e}_{\mathrm{RCM},i}
\left(
\mathbf{X}_{\mathrm{RCM}}
\right)
\right)
\right\}_{i=1}^{L}
\right).
\end{equation}

\item \textbf{Decision-level fusion:}
\begin{equation}
\begin{gathered}
\hat{\mathcal{Y}}_{\mathrm{D},t}
=
w_t\,\hat{\mathcal{Y}}_{\mathrm{S1},t}
+
(1-w_t)\hat{\mathcal{Y}}_{\mathrm{RCM},t}
\\[2mm]
t \in
\{\mathrm{SIC},\mathrm{SOD},\mathrm{FLOE}\}.
\end{gathered}
\end{equation}
\end{enumerate}

\subsection{Model Implementation}

The network designed in this study is based on the U-Net architecture \cite{Ronneberger2015}, adapted for multi-task sea ice mapping as illustrated in Fig.~\ref{fig:model_architectures}. The U-shaped encoder--decoder design enables the network to learn both high-level contextual representations and fine-scale spatial details. Skip connections transfer encoder features directly to the corresponding decoder stages, preserving spatial information during reconstruction and helping address the vanishing-gradient problem. This architecture is particularly suitable for sea ice mapping, where annotated data are limited, as U-Net has shown strong performance in learning from relatively small datasets while maintaining generalization to complex remote-sensing scenes \cite{Radhakrishnan2021, Stokholm2022, deLoe2025, Jamali2024, Ronneberger2015}.

Details of the U-Net architecture used in the experiments are summarized in Table~\ref{tab:unet_architecture}. The encoder consists of five stages, while the decoder progressively restores the spatial resolution through four upsampling stages with gated skip connections. Gated skip connections are adopted to adaptively control the contribution of encoder features during feature reconstruction \cite{Jiang2021}. Unlike standard U-Net skip connections, which directly pass encoder features to the decoder, the gating operation modulates the skip features before they are combined with the decoder features, allowing the network to preserve useful spatial details while reducing the influence of noisy or less relevant low-level SAR responses. This is particularly useful for SAR-based sea ice mapping, where sensor-dependent noise patterns and ambiguous backscatter signatures can affect dense pixel-wise prediction.

As shown in Table~\ref{tab:unet_architecture}, C, P, and T denote convolution, max-pooling, and transposed convolution, respectively. Each $3 \times 3$ convolutional layer is followed by batch normalization and a ReLU activation function. The final decoder feature map is shared by three task-specific prediction heads. The SIC head is implemented as a regression head to produce continuous sea ice concentration estimates, whereas the SOD and FLOE heads are implemented as classification heads for semantic segmentation. Separate $1 \times 1$ convolutional layers are used to generate the final pixel-wise prediction map for each task. Each task-specific output layer uses a number of filters corresponding to the target output dimensionality, denoted as $N_t$. The SIC regression head produces one continuous estimate per pixel ($N_{\mathrm{SIC}}=1$), although the reference SIC labels comprise 11 discrete concentration levels. In contrast, the SOD and FLOE classification heads produce class scores for six and seven categories, respectively ($N_{\mathrm{SOD}}=6$ and $N_{\mathrm{FLOE}}=7$).

\begin{table}[!t]
\centering
\caption{U-Net architecture used in the experiments. C, P, and T denote convolution, max-pooling, and transposed convolution, respectively. $N_t$ denotes the number of output channels for task $t$.}
\label{tab:unet_architecture}
\footnotesize
\begin{tabular}{ll}
\hline
\textbf{Encoder} & \textbf{Decoder} \\
\hline
C$(3 \times 3, 96) \times 2$        & T$(2 \times 2, 384)$ \\
P$(2 \times 2)$                     & C$(3 \times 3, 384) \times 2$ \\
\hline
C$(3 \times 3, 192) \times 3$       & T$(2 \times 2, 384)$ \\
P$(2 \times 2)$                     & C$(3 \times 3, 384) \times 2$ \\
\hline
C$(3 \times 3, 384) \times 3$       & T$(2 \times 2, 192)$ \\
P$(2 \times 2)$                     & C$(3 \times 3, 192) \times 2$ \\
\hline
C$(3 \times 3, 384) \times 4$       & T$(2 \times 2, 96)$ \\
P$(2 \times 2)$                     & C$(3 \times 3, 96) \times 2$ \\
\hline
C$(3 \times 3, 384) \times 4$       & C$(1 \times 1, N_t)$ \\
\hline
\end{tabular}
\end{table}

\subsection{Single-Sensor Baselines}
Two single-sensor baselines are first established to quantify the independent contribution of each SAR platform and provide reference points for evaluating the fusion strategies.
\subsubsection{Sentinel-1-Only Model}
The Sentinel-1-only model uses the dual-polarization Sentinel-1 HH/HV channels as input. This model provides the primary reference baseline, allowing a controlled comparison between single-sensor and fused configurations.
\subsubsection{RCM-Only Model}
The RCM-only model uses the co-registered dual-polarization RCM HH/HV channels as input. This baseline evaluates the standalone capability of RCM imagery for predicting SIC, SOD, and FLOE after spatial alignment with the Sentinel-1 reference grid.

\subsection{Cross-Sensor Fusion Strategies}
The proposed benchmark evaluates three cross-sensor fusion strategies, illustrated in Fig.~\ref{fig:model_architectures}: input-level fusion, feature-level fusion, and decision-level fusion.
\subsubsection{Input-Level Fusion}
Input-level fusion combines the dual-polarization HH/HV channels from Sentinel-1 and RCM observations before feature extraction, forming a four-channel SAR input that is passed to a single encoder--decoder network. This strategy enables the initial convolutional layers to learn shared low-level radiometric and textural patterns directly from the stacked cross-sensor input, but it does not explicitly separate sensor-specific feature pathways. This experiment evaluates whether simple channel concatenation is sufficient to exploit complementary information between Sentinel-1 and RCM, and provides a direct comparison with the more structured feature-level fusion strategy.

\subsubsection{Feature-Level Fusion}
Feature-level fusion uses two parallel encoder branches to learn sensor-specific representations separately from Sentinel-1 and RCM observations before combining them. Each encoder receives the HH/HV channels from one SAR platform and extracts hierarchical feature maps at multiple spatial scales. At each encoder stage, the corresponding Sentinel-1 and RCM feature maps are passed to a fusion block, where they are normalized, concatenated along the channel dimension, and then refined using convolutional layers to produce a fused feature representation. This design allows each sensor to first learn its own radiometric characteristics, noise patterns, and acquisition geometries, while still enabling cross-sensor information exchange before decoding. During decoding, each fused feature map is injected into the corresponding decoder stage to refine the features obtained from the upsampled decoder output and the gated encoder skip connection. The decoder then progressively restores spatial resolution and generates the three output maps for SIC, SOD, and FLOE. Compared with input-level fusion, feature-level fusion provides a more structured mechanism for integrating cross-sensor representations while preserving sensor-specific feature-extraction pathways.

\subsubsection{Decision-Level Fusion}
Decision-level fusion combines the outputs of independently trained Sentinel-1-only and RCM-only models. Unlike input- and feature-level fusion, the sensor streams do not interact during feature extraction. Instead, the two trained models are frozen, and during inference, their independent prediction maps for SIC, SOD, and FLOE are combined at the decision level using validation-selected task-specific weights. For each random seed, the Sentinel-1-only and RCM-only checkpoints selected according to the validation combined score are used to generate all three task outputs. For SIC, the continuous regression maps from the Sentinel-1 and RCM models are combined using a weighted average, expressed as follows:
\begin{equation}
\hat{Y}^{\mathrm{Fused}}_{\mathrm{SIC}}
=
w^{\mathrm{S1}}_{\mathrm{SIC}} \cdot \hat{Y}^{\mathrm{S1}}_{\mathrm{SIC}}
+
w^{\mathrm{RCM}}_{\mathrm{SIC}} \cdot \hat{Y}^{\mathrm{RCM}}_{\mathrm{SIC}} .
\end{equation} For SOD and FLOE, the class-probability maps from the two models are first combined using task-specific weights, and the final class label is then selected as the class with the highest fused probability, expressed as follows:
\begin{equation}
P^{\mathrm{Fused}}_{\mathrm{SOD}}
=
w^{\mathrm{S1}}_{\mathrm{SOD}} \cdot P^{\mathrm{S1}}_{\mathrm{SOD}}
+
w^{\mathrm{RCM}}_{\mathrm{SOD}} \cdot P^{\mathrm{RCM}}_{\mathrm{SOD}} .
\end{equation}
\begin{equation}
Y^{\mathrm{Fused}}_{\mathrm{SOD}}
=
\arg\max P^{\mathrm{Fused}}_{\mathrm{SOD}} .
\end{equation}

\begin{equation}
P^{\mathrm{Fused}}_{\mathrm{FLOE}}
=
w^{\mathrm{S1}}_{\mathrm{FLOE}} \cdot P^{\mathrm{S1}}_{\mathrm{FLOE}}
+
w^{\mathrm{RCM}}_{\mathrm{FLOE}} \cdot P^{\mathrm{RCM}}_{\mathrm{FLOE}} .
\end{equation} 
\begin{equation}
Y^{\mathrm{Fused}}_{\mathrm{FLOE}}
=
\arg\max P^{\mathrm{Fused}}_{\mathrm{FLOE}} .
\end{equation} The fusion weights are selected independently for each task using a grid search over Sentinel-1 weights ranging from 0 to 1 in increments of 0.1 (11 candidate values), based on the validation-set performance of the independently trained single-sensor models. For SIC, the weight pair with the highest validation $R^{2}$ is selected, while for SOD and FLOE, the weight pairs with the highest validation F1-score are used. This allows each sea ice parameter to use the combination of Sentinel-1 and RCM contributions that provides the best validation performance. Weight selection is performed separately for each cross-validation fold and random seed, and the resulting weights are applied to the corresponding held-out test set. This yields 15 sets of task-specific fusion weights across five folds and three random seeds, whose mean and standard deviation are summarized in Table~\ref{tab:decision_fusion_weights}.

\begin{table}[!t]
\centering
\caption{Validation-selected decision-level fusion weights
across five cross-validation folds and three random seeds.
Values are reported as mean $\pm$ standard deviation
across 15 runs.}
\label{tab:decision_fusion_weights}
\footnotesize
\renewcommand{\arraystretch}{1.5}
\begin{tabular}{lcc}
\hline
\textbf{Task} &
\textbf{Sentinel-1 Weights} &
\textbf{RCM Weights} \\
\hline
SIC  & $0.713 \pm 0.203$ & $0.287 \pm 0.203$ \\
SOD  & $0.480 \pm 0.416$ & $0.520 \pm 0.416$ \\
FLOE & $0.553 \pm 0.119$ & $0.447 \pm 0.119$ \\
\hline
\end{tabular}
\end{table}

Overall, the task-specific weights indicate that the validation-optimal sensor contribution differs across sea ice parameters. For each fold and random seed, the corresponding validation-selected weights are applied separately to the SIC, SOD, and FLOE outputs, and the final decision-level results are reported as the mean and standard deviation across the five folds and three seeds. For SIC, the selected weights favor Sentinel-1, with an average weight of 0.713 compared with 0.287 for RCM. For FLOE, the weights are more balanced, with relatively low variability across runs. In contrast, SOD exhibits substantial variability in the selected weights, suggesting that the validation-optimal sensor contribution is sensitive to the cross-validation partition and random seed. This decision-level experiment evaluates whether combining predictions from the two SAR platforms can improve performance without requiring joint feature learning.

\subsection{Experimental Configurations}
\label{sec:experimental_configurations}
All models are optimized under the same multi-task learning framework. The specifications of model training are detailed in Table~\ref{tab:training_settings}. 

To evaluate model generalization and reduce the dependence of the results on a single data partition, five-fold cross-validation was performed at the scene level using the 221 paired SAR scenes. The scenes were randomly divided into five folds. In each iteration, four folds (approximately 80\% of the scenes) were used for model development, while the remaining fold (approximately 20\%) was held out for testing. Within the development set, 90\% of the scenes were used for training and 10\% for validation. This procedure was repeated across all five folds, ensuring that each scene was included in the test set exactly once. For each fold, three independent training runs were conducted using random seeds 42, 123, and 3407, resulting in 15 independent training runs for each trainable configuration. Validation and test performance were averaged across the three random seeds within each fold, and the final results are reported as the mean and standard deviation across the five fold-level averages.

To expose the models to different effective spatial contexts, we perform multi-scale training by extracting patches generated with downsampling factors ranging from $2\times$ to $10\times$. During training, $384 \times 384$ image patches are randomly sampled from these multi-scale datasets with a batch size of 16. For data augmentation, the training pipeline includes random resizing, cropping, and horizontal flipping. The models are trained using the AdamW optimizer with an initial learning rate of $3 \times 10^{-4}$ and a weight decay of $10^{-3}$. Cosine annealing~\cite{Loshchilov2016} is used as the learning-rate schedule, gradually reducing the learning rate according to a cosine decay function until it reaches a minimum value. Training is performed for 60,000 iterations, and validation is conducted every 1,000 iterations. 
After exploring different loss-function configurations, we found that formulating SIC as a regression task with mean squared error (MSE) loss and treating SOD and FLOE as semantic classification tasks with cross-entropy (CE) loss produce the highest validation performance. The MSE loss function is defined as follows:
\begin{equation}
\mathcal{L}_{\mathrm{SIC}}
=
\frac{1}{N}
\sum_{i=1}^{N}
\left(
\hat{y}^{\mathrm{SIC}}_{i}
-
y^{\mathrm{SIC}}_{i}
\right)^2 ,
\end{equation} The CE loss function is defined as follows:
\begin{equation}
\mathcal{L}_{t}
=
-
\frac{1}{N}
\sum_{i=1}^{N}
\log
\left(
\hat{p}^{t}_{i,y_i}
\right),
\quad
t \in \{\mathrm{SOD}, \mathrm{FLOE}\},
\end{equation} where $N$ is the number of valid pixels and $\hat{p}^{t}_{i,y_i}$ is the predicted probability of the correct class $y_i$ at pixel $i$. Considering that the magnitude of the MSE loss used for SIC is higher than that of the CE losses used for SOD and FLOE, larger weights are assigned to the CE losses when computing the total loss. The larger CE weights were determined empirically to balance the contributions of the three tasks during training. The total loss is therefore defined as follows:
\begin{equation}
\mathcal{L}_{\mathrm{total}}
=
\mathcal{L}_{\mathrm{SIC}}
+
3\mathcal{L}_{\mathrm{SOD}}
+
3\mathcal{L}_{\mathrm{FLOE}} .
\end{equation}

To evaluate model generalization during training, a separate validation subset is used within each cross-validation fold. At each validation interval, the evaluation metrics described in Section~\ref{sec:evaluation_metrics} are computed on the corresponding validation set, and the combined score is calculated using Eq.~\eqref{eq:combined_score}. Model checkpoints are saved based on the combined score and the task-specific validation metrics. The checkpoints selected according to the validation combined score are then used to generate predictions for the test set. For decision-level fusion, the same selected Sentinel-1-only and RCM-only checkpoints are used to generate the SIC, SOD, and FLOE predictions before applying the corresponding task-specific fusion weights.

All experiments were implemented using the OpenMMLab framework and conducted on the Rorqual cluster of the Digital Research Alliance of Canada, Canada’s national high-performance computing system. Each experiment was run on one NVIDIA H100 80GB HBM3 GPU with six CPU cores and 80~GB of system memory, using PyTorch 2.12.0. On average, training each model required approximately 30 hours.

\begin{table}[!t]
\centering
\caption{Training and inference settings used in the experiments.}
\label{tab:training_settings}
\footnotesize
\renewcommand{\arraystretch}{1.3}
\begin{tabular}{p{0.42\linewidth}p{0.48\linewidth}}
\hline
\textbf{Setting} & \textbf{Value} \\
\hline
Optimizer & AdamW \\
Initial learning rate & $3 \times 10^{-4}$ \\
Weight decay & $1 \times 10^{-3}$ \\
Learning-rate scheduler & Cosine annealing \\
Minimum learning rate & $1 \times 10^{-6}$ \\
Iterations & 60,000 \\
Validation interval & Every 1,000 iterations \\
Cross-validation & 5 folds \\
Independent runs & 3 (seeds 42, 123, and 3407) \\
Batch size & 16 \\
Patch size & $384 \times 384$ \\
Downsampling ratio & $2\times$--$10\times$ \\
Data augmentation & Random resize, crop, flip \\
SIC loss & MSE loss \\
SOD/FLOE loss & Cross-entropy loss \\
Total loss calculation & SIC $\times 1$ + SOD $\times 3$ + FLOE $\times 3$ \\
\hline
\end{tabular}
\end{table}

\subsection{Evaluation Metrics}
\label{sec:evaluation_metrics}

The models are evaluated separately for the three sea ice parameters. SIC is assessed using the coefficient of determination, $R^{2}$, which measures the agreement between predicted and reference SIC values. The $R^{2}$ score is defined as follows:
\begin{equation}
R^{2}
=
1 -
\frac{
\sum_{i=1}^{N}
\left(
y_i - \hat{y}_i
\right)^2
}{
\sum_{i=1}^{N}
\left(
y_i - \bar{y}
\right)^2
},
\end{equation}
where $y_i$ and $\hat{y}_i$ are the reference and predicted SIC values at pixel $i$, respectively, and $\bar{y}$ is the mean reference SIC value. $N$ denotes the number of valid pixels.
SOD and FLOE are evaluated using F1-score to account for class imbalance among sea ice categories. For each class, the F1-score is computed from precision and recall as follows:
\begin{equation}
F_{1}
=
2
\frac{
\mathrm{Precision} \cdot \mathrm{Recall}
}{
\mathrm{Precision} + \mathrm{Recall}
},
\end{equation}
where
\begin{equation}
\mathrm{Precision}
=
\frac{TP}{TP+FP},
\qquad
\mathrm{Recall}
=
\frac{TP}{TP+FN}.
\end{equation} Here, $TP$, $FP$, and $FN$ denote the number of true positives, false positives, and false negatives, respectively. For SOD and FLOE, the overall task-level F1-score is computed as the support-weighted average of the class-wise F1-scores, with each class weighted according to its number of reference-label pixels.

\begin{equation}
F_{1,\mathrm{weighted}}
=
\frac{\sum_{c=1}^{C} n_c F_{1,c}}
{\sum_{c=1}^{C} n_c},
\end{equation}
where $n_c$ denotes the number of reference-label pixels belonging to class $c$. Per-class F1-scores are additionally reported to examine performance on individual sea ice categories. In addition, overall accuracy (aAcc) is reported for each task as a complementary measure of pixel-wise agreement. It is computed as the ratio of correctly predicted valid pixels to the total number of valid pixels:
\begin{equation}
\mathrm{aAcc}
=
\frac{
\sum_{c=1}^{C} TP_c
}{
N
},
\end{equation}
where $TP_c$ denotes the number of correctly predicted pixels for class $c$, $C$ is the number of classes, and $N$ is the total number of valid pixels. 

For SIC, the network produces continuous regression estimates, while the reference labels are represented by 11 discrete concentration levels from 0 to 10, corresponding to 0\%--100\% ice concentration in 10\% increments. Following the discretization approach used in previous SIC regression studies~\cite{deLoe2025}, the continuous predictions are rounded to the nearest concentration level and restricted to the valid range of 0--10 for computing SIC aAcc, per-bin accuracy, and the confusion matrix. The resulting discrete predictions are then compared with the corresponding reference SIC bins.

The three task-specific scores are aggregated into a single combined score according to Eq.~\eqref{eq:combined_score}, providing an overall measure of model performance across the multi-task sea ice mapping benchmark. This enables consistent comparison among the evaluated single-sensor and fusion-based models. Following the evaluation protocol of the AutoICE challenge \cite{Stokholm2024}, the three task-specific metrics are combined using weights of 0.4, 0.4, and 0.2 for SIC, SOD, and FLOE, respectively. We adopt the same weighting scheme to maintain consistency with the established AI4Arctic benchmark. The combined score is computed as:
\begin{equation}
    \mathrm{Combined\ Score}
    =
    0.4R^{2}_{\mathrm{SIC}}
    +
    0.4F_{1,\mathrm{SOD}}
    +
    0.2F_{1,\mathrm{FLOE}}.
    \label{eq:combined_score}
\end{equation}
SIC and SOD are assigned larger weights than FLOE, consistent with ice-charting experts’ assessment of their greater importance to ice services and end users \cite{Stokholm2024}.
To further interpret model behavior, qualitative prediction maps are generated for each task. For SIC, per-bin accuracy and the binned confusion matrix are analyzed to evaluate performance across different concentration levels. For SOD and FLOE, confusion matrices, per-class accuracy, and per-class F1-scores are used to identify class-specific errors and assess which sea ice categories benefit most from cross-sensor SAR fusion.

\section{Results and Discussion}
\label{sec:results}
\subsection{Relative Model Performance}
Tables~\ref{tab:validation_results} and~\ref{tab:main_results} summarize the validation- and test-set performance of single-sensor baselines and cross-sensor fusion strategies using five-fold scene-level cross-validation. Results are reported as the mean and standard deviation across five folds, with performance averaged over three independent training runs within each fold.

The validation results in Table~\ref{tab:validation_results} show that decision-level fusion achieves the highest combined score of $87.30 \pm 4.18\%$, followed by feature-level fusion at $86.49 \pm 3.07\%$. The task-specific results, however, show that no single configuration performs best across all metrics. Sentinel-1-only achieves the highest SIC aAcc of $66.06 \pm 7.17\%$, while decision-level fusion obtains the highest SIC $R^2$ score of $92.31 \pm 1.62\%$, SOD aAcc of $84.13 \pm 6.37\%$, and SOD F1-score of $82.74 \pm 7.07\%$. Feature-level fusion achieves the highest FLOE aAcc of $86.13 \pm 3.37\%$ and FLOE F1-score of $86.81 \pm 3.57\%$. The RCM-only model also achieves a slightly higher SOD F1-score than the Sentinel-1-only model, at $80.69 \pm 5.73\%$ compared with $80.65 \pm 7.40\%$. Overall, the validation results indicate that both single-sensor and cross-sensor configurations exhibit task-dependent behavior, with fusion strategies generally achieving better results than single-sensor configurations on the validation set. For decision-level fusion, the validation results reflect weights selected on the validation set, while generalization is evaluated on the held-out test fold.

The test-set results in Table~\ref{tab:main_results} show a different pattern. Input-level fusion achieves the highest mean combined score of $85.81 \pm 2.21\%$, but this is only marginally higher than the Sentinel-1-only baseline at $85.48 \pm 2.00\%$. Input-level fusion achieves the highest SIC aAcc of $66.89 \pm 3.89\%$ and SIC $R^{2}$ of $91.86 \pm 0.61\%$, while feature-level fusion achieves the highest SOD aAcc of $82.75 \pm 3.18\%$ and SOD F1-score of $80.80 \pm 3.71\%$. Decision-level fusion achieves the highest FLOE aAcc of $85.16 \pm 3.48\%$ and FLOE F1-score of $84.58 \pm 4.39\%$, indicating that task-specific output weighting can provide modest benefits for certain prediction tasks. Overall, the improvements achieved by fusion strategies remain small compared with the Sentinel-1-only baseline.

A notable difference between the validation and test results is observed for the fusion models. Although decision-level fusion achieves the highest combined score on the validation set at $87.30 \pm 4.18\%$, its test combined score decreases to $85.45 \pm 2.04\%$, remaining comparable to the Sentinel-1-only baseline at $85.48 \pm 2.00\%$. In contrast, input- and feature-level fusion achieve the highest test combined scores of $85.81 \pm 2.21\%$ and $85.65 \pm 2.88\%$, respectively, representing only marginal improvements over Sentinel-1-only. These differences may partly reflect variations in class composition and pixel-level representation between the validation and test subsets, particularly for less frequent SIC, SOD, and FLOE categories. Despite achieving the highest validation score, decision-level fusion does not demonstrate a clear advantage on the test set. These results indicate that cross-sensor fusion does not automatically lead to substantial improvements in generalization and that its effectiveness depends on both the fusion strategy and the prediction task.

\begin{table*}[!t]
\centering
\renewcommand{\arraystretch}{1.2}
\caption{Quantitative validation-set comparison of single-sensor baselines and cross-sensor fusion strategies using five-fold cross-validation. The combined score summarizes SIC $R^{2}$, SOD F1-score, and FLOE F1-score. Values are reported as the mean $\pm$ standard deviation across five folds, with results averaged over three independent runs per fold.}
\label{tab:validation_results}
\footnotesize
\begin{tabular}{lccccccc}
\hline
\multirow{2}{*}{\textbf{Model}}
& \multirow{2}{*}{\textbf{Combined Score [\%]}}
& \multicolumn{2}{c}{\textbf{SIC}}
& \multicolumn{2}{c}{\textbf{SOD}}
& \multicolumn{2}{c}{\textbf{FLOE}} \\
\cline{3-8}
&
& \textbf{aAcc [\%]} & \boldmath{$R^{2}$ \textbf{[\%]}}
& \textbf{aAcc [\%]} & \boldmath{$F_{1}$ \textbf{[\%]}}
& \textbf{aAcc [\%]} & \boldmath{$F_{1}$ \textbf{[\%]}} \\
\hline
Sentinel-1-only
& 85.86 $\pm$ 4.46
& \textbf{66.06 $\pm$ 7.17} & 91.17 $\pm$ 2.22
& 82.00 $\pm$ 6.41 & 80.65 $\pm$ 7.40
& 84.75 $\pm$ 4.76 & 85.69 $\pm$ 4.39 \\

RCM-only
& 84.26 $\pm$ 3.04
& 61.04 $\pm$ 5.35 & 87.38 $\pm$ 1.99
& 81.56 $\pm$ 5.48 & 80.69 $\pm$ 5.73
& 84.04 $\pm$ 4.72 & 85.18 $\pm$ 4.20 \\

Input-level fusion
& 85.96 $\pm$ 3.89
& 65.67 $\pm$ 5.90 & 90.96 $\pm$ 2.24
& 81.48 $\pm$ 5.55 & 81.29 $\pm$ 6.18
& 84.17 $\pm$ 4.69 & 85.28 $\pm$ 4.42 \\

Feature-level fusion
& 86.49 $\pm$ 3.07
& 63.63 $\pm$ 6.19 & 90.64 $\pm$ 1.35
& 82.87 $\pm$ 5.07 & 82.19 $\pm$ 5.17
& \textbf{86.13 $\pm$ 3.37} & \textbf{86.81 $\pm$ 3.57} \\

Decision-level fusion
& \textbf{87.30 $\pm$ 4.18}
& 61.96 $\pm$ 6.92 & \textbf{92.31 $\pm$ 1.62}
& \textbf{84.13 $\pm$ 6.37} & \textbf{82.74 $\pm$ 7.07}
& 86.07 $\pm$ 4.72 & 86.42 $\pm$ 4.52 \\
\hline
\end{tabular}
\end{table*}

\begin{table*}[!t]
\centering
\renewcommand{\arraystretch}{1.2}
\caption{Quantitative test-set comparison of single-sensor baselines and cross-sensor fusion strategies using five-fold cross-validation. The combined score summarizes SIC $R^{2}$, SOD F1-score, and FLOE F1-score. Values are reported as the mean $\pm$ standard deviation across five folds, with results averaged over three independent runs per fold.}
\label{tab:main_results}
\footnotesize
\begin{tabular}{lccccccc}
\hline
\multirow{2}{*}{\textbf{Model}}
& \multirow{2}{*}{\textbf{Combined Score [\%]}}
& \multicolumn{2}{c}{\textbf{SIC}}
& \multicolumn{2}{c}{\textbf{SOD}}
& \multicolumn{2}{c}{\textbf{FLOE}} \\
\cline{3-8}
&
& \textbf{aAcc [\%]} & \boldmath{$R^{2}$ \textbf{[\%]}}
& \textbf{aAcc [\%]} & \boldmath{$F_{1}$ \textbf{[\%]}}
& \textbf{aAcc [\%]} & \boldmath{$F_{1}$ \textbf{[\%]}} \\
\hline
Sentinel-1-only
& 85.48 $\pm$ 2.00
& 65.56 $\pm$ 3.92 & 90.85 $\pm$ 0.87
& 82.44 $\pm$ 2.56 & 80.70 $\pm$ 3.05
& 84.23 $\pm$ 3.43 & 84.29 $\pm$ 4.14 \\

RCM-only
& 82.56 $\pm$ 3.38
& 59.55 $\pm$ 6.95 & 86.00 $\pm$ 2.40
& 81.79 $\pm$ 2.66 & 79.04 $\pm$ 4.02
& 83.13 $\pm$ 4.21 & 82.72 $\pm$ 5.35 \\

Input-level fusion
& \textbf{85.81 $\pm$ 2.21}
& \textbf{66.89 $\pm$ 3.89} & \textbf{91.86 $\pm$ 0.61}
& 82.10 $\pm$ 2.71 & 80.50 $\pm$ 3.34
& 84.20 $\pm$ 2.52 & 84.34 $\pm$ 3.75 \\

Feature-level fusion
& 85.65 $\pm$ 2.88
& 66.33 $\pm$ 4.07 & 91.04 $\pm$ 1.29
& \textbf{82.75 $\pm$ 3.18} & \textbf{80.80 $\pm$ 3.71}
& 84.93 $\pm$ 4.23 & 84.56 $\pm$ 4.92 \\

Decision-level fusion
& 85.45 $\pm$ 2.04
& 62.99 $\pm$ 3.53 & 90.96 $\pm$ 0.79
& 82.69 $\pm$ 2.19 & 80.38 $\pm$ 2.98
& \textbf{85.16 $\pm$ 3.48} & \textbf{84.58 $\pm$ 4.39} \\
\hline
\end{tabular}
\end{table*}

\subsection{Comparison of Predicted Sea Ice Maps}
\begin{figure*}[!t]
\centering
\includegraphics[width=0.80\textwidth]{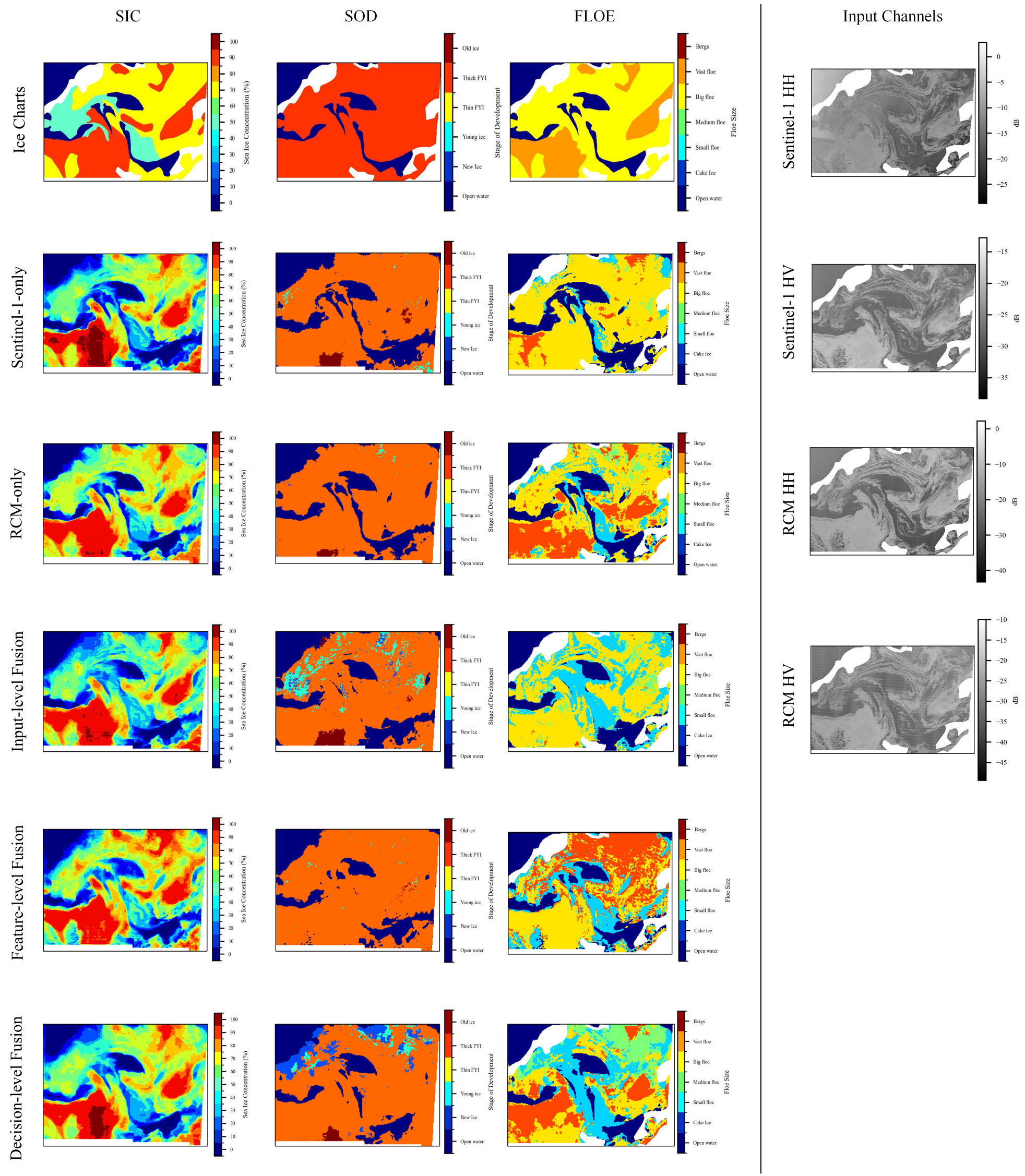}
\caption{Qualitative comparison of SIC, SOD, and FLOE predictions for an example test scene, acquired on April 24, 2020, near southern Greenland in the Davis Strait region. The figure shows the Sentinel-1 and RCM HH/HV input channels, the ice-chart reference labels, and the prediction maps from the Sentinel-1-only, RCM-only, input-level fusion, feature-level fusion, and decision-level fusion models. RADARSAT Constellation Mission Imagery
\textcopyright\ Government of Canada (2019–2021).
RADARSAT is an official mark of the Canadian Space Agency.}
\label{fig:qualitative_results}
\end{figure*}

Fig.~\ref{fig:qualitative_results} provides a qualitative comparison of the predicted SIC, SOD, and FLOE maps for an example test scene acquired on April 24, 2020, near southern Greenland in the Davis Strait region. The figure includes the ice-chart-derived reference labels, predictions from all evaluated models, and the corresponding Sentinel-1 and RCM HH/HV input channels, which are also shown to provide visual context for the cross-sensor SAR inputs. The qualitative results provide additional context for the quantitative trends in Table~\ref{tab:main_results}. All evaluated models capture the main spatial patterns of SIC, SOD, and FLOE to varying degrees, but their predictions differ in spatial consistency and class representation.

For SIC, all models capture the main open water region and recover the general transition from low to high sea ice concentration across the ice-covered area. The predicted SIC maps exhibit more gradual spatial transitions than the polygon-based ice-chart labels, as SIC is treated as a regression task. Sentinel-1-only and RCM-only produce broadly similar concentration patterns, while the fusion models exhibit differences in the spatial distribution of intermediate and high concentration values. Decision-level fusion preserves the broad spatial structure of the SIC chart, although local differences remain across the predicted maps.

For SOD, all models identify the dominant Open Water and Thick First-Year Ice regions, while less frequent development stages remain difficult to recover. Sentinel-1-only and RCM-only produce relatively similar maps, with some local inconsistencies near class boundaries. Input-level and decision-level fusion introduce additional fragmented predictions, particularly within the dominant Thick First-Year Ice region. In contrast, feature-level fusion produces a relatively homogeneous SOD map, preserving the main separation between open water and ice-covered regions. However, this spatial consistency does not necessarily indicate better identification of less frequent classes. 

FLOE prediction shows the largest visual differences among the models. Sentinel-1-only and RCM-only capture the general distribution of the dominant floe-size categories but exhibit fragmented predictions, particularly near transitions between Small, Medium, Big, and Vast Floe classes. Input-level fusion produces a similarly fragmented map, while feature-level fusion shows substantial spatial variation and a larger extent of Vast Floe predictions compared with the reference chart. Decision-level fusion better preserves some of the broader floe-size regions, although local inconsistencies and class confusion remain evident. These observations suggest that FLOE prediction is sensitive to the choice of fusion strategy, while incorporating RCM provides limited additional benefits over Sentinel-1 alone.

\begin{figure*}[!t]
\centering
\includegraphics[width=\textwidth]{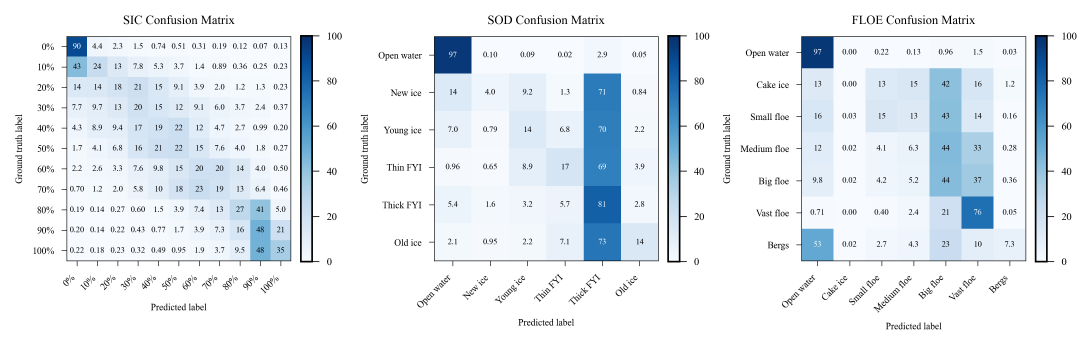}
\caption{Row-normalized confusion matrices of the Sentinel-1-only model for SIC, SOD, and FLOE on the held-out test sets. Confusion-matrix counts were aggregated across five cross-validation folds and three random seeds before normalization. Rows represent reference labels, and columns represent predicted labels.}
\label{fig:s1_only_15_run_confusion_matrices}
\end{figure*}

% \begin{figure*}[!t]
% \centering
% \includegraphics[width=\textwidth]{figures/rcm_only_15_run_confusion_matrices.png}
% \caption {Normalized confusion matrices of the RCM-only model for SIC, SOD, and FLOE on the test set. Rows represent ground-truth labels and columns represent predicted labels.}
% \label{fig:rcm_only_15_run_confusion_matrices}
% \end{figure*}

\subsection{SIC Analysis}
The SIC results show that the relative performance of the evaluated configurations depends on the evaluation metric. As reported in Table~\ref{tab:main_results}, input-level fusion achieves the highest SIC $R^{2}$ of $91.86\pm0.61\%$, compared with $90.85\pm0.87\%$ for Sentinel-1-only. Input-level fusion also achieves the highest SIC aAcc of $66.89\pm3.89\%$, followed by feature-level fusion at $66.33\pm4.07\%$. Both input- and feature-level fusion achieve higher aAcc and $R^{2}$ values than Sentinel-1-only, although the improvements remain modest. These results indicate that incorporating RCM observations can provide limited benefits for both discrete SIC-bin agreement and continuous SIC estimation, without substantially improving overall prediction performance.

\begin{figure*}[!t]
\centering
\includegraphics[width=0.85\textwidth]{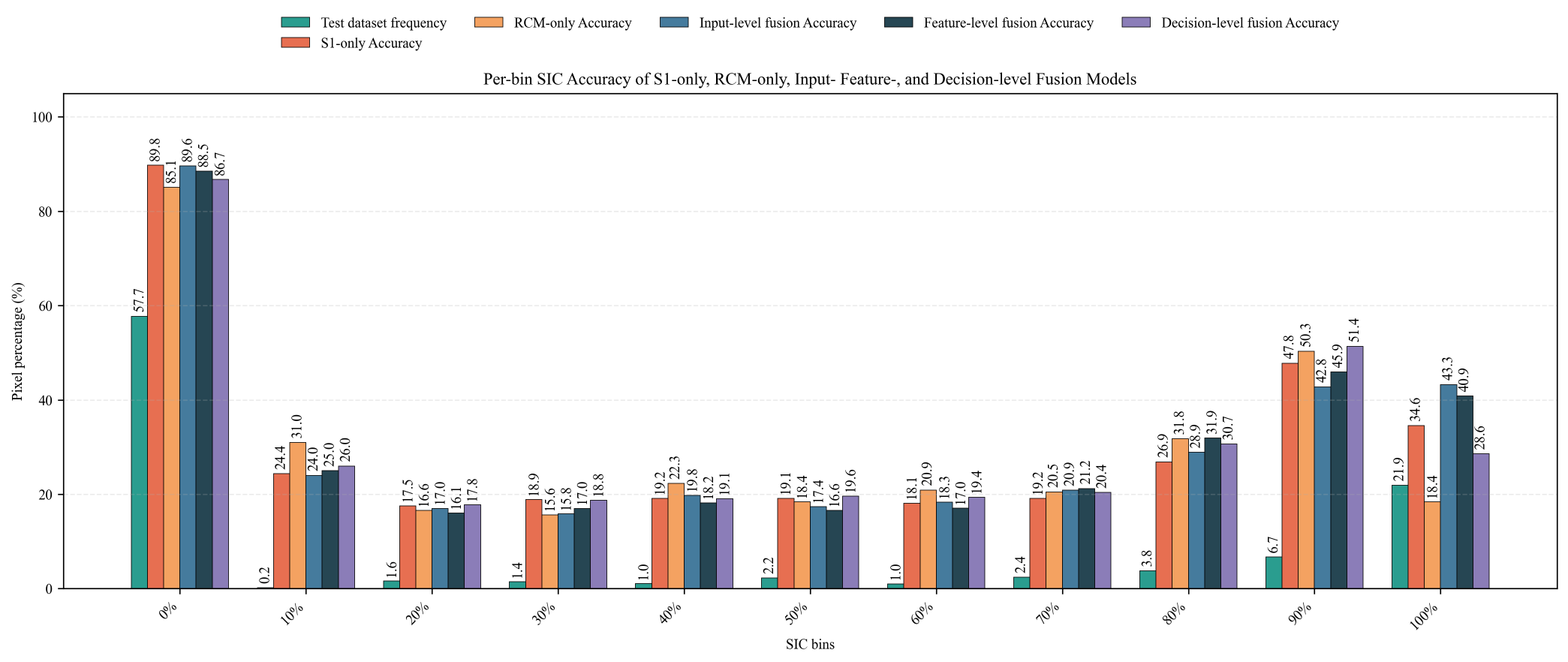}
\caption{Per-bin SIC accuracy of Sentinel-1-only, RCM-only, input-level fusion, feature-level fusion, and decision-level fusion models compared with the representation of each SIC bin in the test dataset.}
\label{fig:sic_per_bin_accuracy}
\end{figure*}

Fig.~\ref{fig:sic_per_bin_accuracy} further compares the per-bin SIC accuracy of selected configurations with the test-set frequency of each SIC level. Performance varies substantially across SIC bins, and no single model consistently achieves the highest accuracy across all concentration levels. The dominant 0\% and 100\% bins generally achieve higher accuracy than many of the less frequent intermediate SIC levels, while several intermediate bins remain difficult for all configurations. The relative performance of the single-sensor and fusion models varies across concentration levels, with different configurations achieving higher accuracy for different SIC bins. Although input-level fusion achieves the highest aggregate SIC aAcc and $R^{2}$, these improvements do not necessarily extend uniformly across all concentration levels. This variation further indicates that the contribution of the two sensors depends on the sea ice concentration regime rather than being uniform across SIC values.

The binned SIC confusion matrix in Fig.~\ref{fig:s1_only_15_run_confusion_matrices} further shows that most remaining errors occur among adjacent or nearby intermediate concentration levels. This reflects the difficulty of estimating mixed ice conditions from SAR imagery. This behavior is related to the limited representation of several intermediate SIC bins, the ambiguity of mixed ice conditions, and the polygon-based nature of the ice-chart-derived labels.

A similar tendency toward SIC overestimation in analyst-assigned reference labels was reported by Cheng et al.~\cite{Cheng2020}, who found that Canadian Ice Service analysts generally estimated higher SIC values than those derived from automated SAR image segmentation. Notably this tendency persisted even when analysts agreed that the segmentation correctly distinguished sea ice from open water. This finding suggests that the apparent underestimation of SIC observed in the present results may partly reflect uncertainty or systematic bias in the ice-chart reference labels, rather than solely model prediction errors.

\subsection{SOD Analysis}

For SOD classification, the test results reveal similar rankings for the F1-score and aAcc. Feature-level fusion achieves the highest SOD F1-score of $80.80\pm3.71\%$, followed closely by Sentinel-1-only at $80.70\pm3.05\%$. Feature-level fusion also achieves the highest SOD aAcc of $82.75\pm3.18\%$, followed by decision-level fusion at $82.69\pm2.19\%$. These results indicate that the relative sensor contribution depends on the model configuration and fusion strategy, although the improvements over Sentinel-1-only remain marginal. Because the reported task-level F1-score is weighted, frequent classes contribute more strongly to the overall F1-score. Per-class results are therefore important for interpreting performance on minority SOD categories.

\begin{figure*}[!t]
\centering
\includegraphics[width=0.85\textwidth]{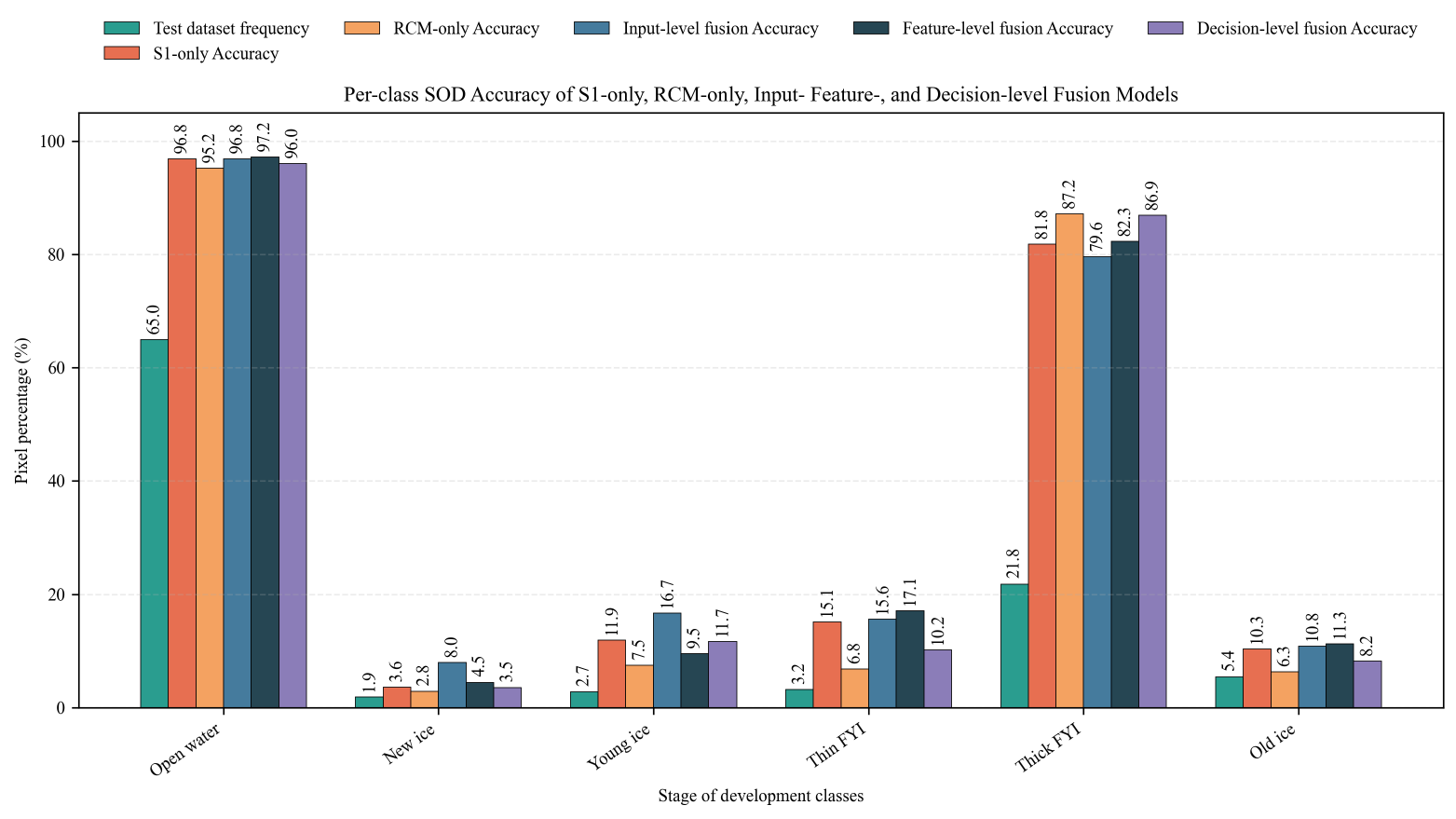}
\caption{Per-class SOD accuracy of Sentinel-1-only, RCM-only, input-level fusion, feature-level fusion, and decision-level fusion models compared with the test-set frequency of each SOD class.}
\label{fig:sod_per_class_accuracy}
\end{figure*}

Fig.~\ref{fig:sod_per_class_accuracy} further illustrates the relationship between class representation and prediction accuracy. Open Water and Thick First-Year Ice dominate the test set and generally achieve higher accuracies than the less frequent SOD categories. In contrast, New Ice, Young Ice, Thin First-Year Ice, and Old Ice remain more difficult to classify across the evaluated models, highlighting the challenges associated with class imbalance in SOD prediction.

The per-class F1-scores in Table~\ref{tab:sod_per_class_f1} show that no single model dominates all SOD categories. Feature-level fusion achieves the highest F1-scores for Open Water and Old Ice, while input-level fusion performs best for New Ice, Young Ice, and Thin First-Year Ice. Decision-level fusion achieves the highest F1-score for Thick First-Year Ice. These class-specific differences demonstrate that the relative usefulness of Sentinel-1 and RCM observations varies across sea ice development stages. However, the improvements for less frequent classes remain limited, with particularly low F1-scores for New Ice, Young Ice, Thin First-Year Ice, and Old Ice across all configurations. These results are consistent with the strong class imbalance in the SOD labels shown in Fig.~\ref{fig:label_distribution}, where these classes represent relatively small fractions of the available pixels.

The Sentinel-1-only confusion matrix in Fig.~\ref{fig:s1_only_15_run_confusion_matrices} further shows that the dominant Open Water and Thick First-Year Ice classes are identified more reliably than the less frequent development stages. In contrast, minority classes exhibit substantial misclassification, with many predictions assigned to the dominant Thick First-Year Ice category. This suggests that the reduced performance is not solely associated with limited class frequency but may also reflect limited separability between certain SOD categories in the learned SAR representations. These errors are consistent with class imbalance, potentially overlapping SAR backscatter characteristics among ice development stages, and the polygon-level nature of the reference labels.

\begin{table*}[!t]
\centering
\renewcommand{\arraystretch}{1.2}
\caption{Per-class SOD F1-scores on the test set for single-sensor baselines and cross-sensor fusion strategies. Values are reported as the mean $\pm$ standard deviation across five folds, with results averaged over three independent runs per fold.}
\label{tab:sod_per_class_f1}
\footnotesize
\begin{tabular}{lcccccc}
\hline
\textbf{Model}
& \textbf{Open Water [\%]}
& \textbf{New Ice [\%]}
& \textbf{Young Ice [\%]}
& \textbf{Thin FYI [\%]}
& \textbf{Thick FYI [\%]}
& \textbf{Old Ice [\%]} \\
\hline

Sentinel-1-only
& 97.04 $\pm$ 0.43
& 5.17 $\pm$ 3.44
& 14.02 $\pm$ 12.63
& 17.68 $\pm$ 15.72
& 69.77 $\pm$ 2.47
& 15.76 $\pm$ 12.03 \\

RCM-only
& 95.78 $\pm$ 0.56
& 3.78 $\pm$ 2.73
& 9.54 $\pm$ 6.39
& 10.62 $\pm$ 9.43
& 69.78 $\pm$ 3.93
& 9.72 $\pm$ 15.25 \\

Input-level fusion
& 96.88 $\pm$ 0.51
& \textbf{7.44 $\pm$ 7.67}
& \textbf{16.91 $\pm$ 12.07}
& \textbf{19.42 $\pm$ 16.46}
& 68.89 $\pm$ 1.89
& 15.43 $\pm$ 14.82 \\

Feature-level fusion
& \textbf{97.14 $\pm$ 0.87}
& 5.52 $\pm$ 3.21
& 11.88 $\pm$ 4.68
& 18.20 $\pm$ 10.29
& 70.32 $\pm$ 4.18
& \textbf{16.88 $\pm$ 12.95} \\

Decision-level fusion
& 96.63 $\pm$ 0.25
& 4.71 $\pm$ 2.98
& 14.10 $\pm$ 9.10
& 14.26 $\pm$ 10.07
& \textbf{70.84 $\pm$ 3.09}
& 12.97 $\pm$ 14.68 \\
\hline
\end{tabular}
\end{table*}

\subsection{FLOE Analysis}
\label{subsec:floe_analysis}
For FLOE prediction, Sentinel-1-only and decision-level fusion both achieve competitive F1-scores of $84.29\pm4.14\%$ and $84.58\pm4.39\%$, respectively. Decision-level fusion also achieves the highest FLOE aAcc of $85.16\pm3.48\%$, as reported in Table~\ref{tab:main_results}. Although decision-level fusion achieves the highest values for both metrics, its improvements over Sentinel-1-only remain marginal. The high aggregate F1-scores and aAcc values are strongly influenced by the dominance of Open Water and Vast Floe, highlighting the importance of examining per-class performance.

\begin{figure*}[!t]
\centering
\includegraphics[width=0.85\textwidth]{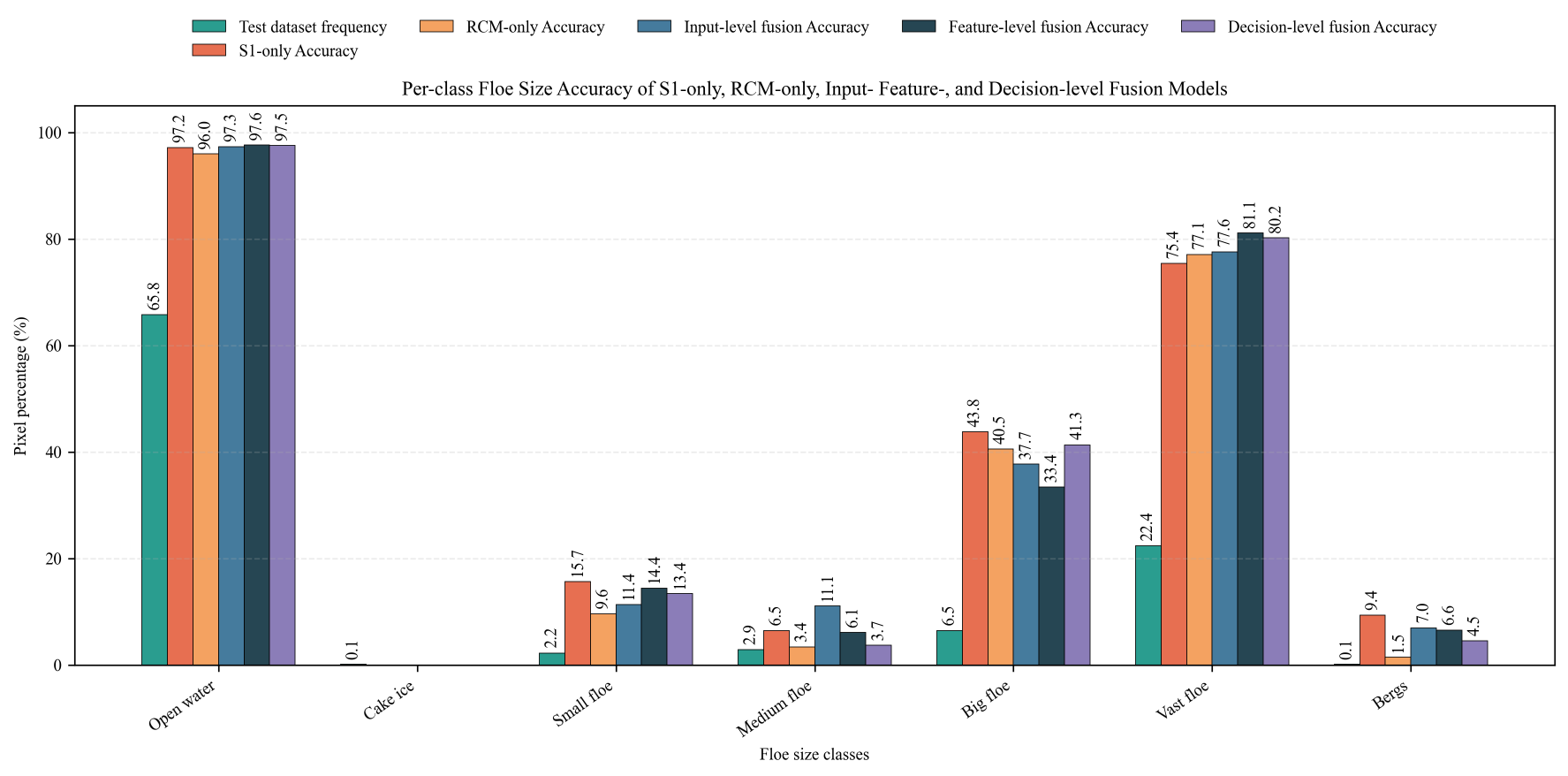}
\caption{Per-class FLOE accuracy of Sentinel-1-only, RCM-only, input-level fusion, feature-level fusion, and decision-level fusion models compared with the test-set frequency of each FLOE class.}
\label{fig:floe_per_class_accuracy}
\end{figure*}

Fig.~\ref{fig:floe_per_class_accuracy} shows that Open Water and Vast Floe dominate the test set and generally achieve substantially higher accuracies than most minority floe-size categories. In contrast, less frequent classes such as Cake Ice, Small Floe, Medium Floe, and Bergs exhibit much lower accuracies, although their relative performance varies among the sensor configurations.

The per-class F1-scores in Table~\ref{tab:floe_per_class_f1} further demonstrate this model- and class-dependent behavior. Feature-level fusion achieves the highest F1-scores for Open Water and Cake Ice, although the F1-score for Cake Ice remains nearly zero. Sentinel-1-only performs best for Small Floe and Bergs, while input-level fusion achieves the highest F1-scores for Medium Floe and Big Floe. Decision-level fusion achieves the highest F1-score for Vast Floe. Thus, no single sensor or fusion strategy consistently dominates across the FLOE categories, and the improvements for minority classes remain limited.

The FLOE confusion matrix in Fig.~\ref{fig:s1_only_15_run_confusion_matrices} shows that Sentinel-1-only identifies the dominant Open Water and Vast Floe classes more reliably than the less frequent floe-size categories. Most errors occur among the minority floe-size classes, particularly Small Floe, Medium Floe, Big Floe, and Bergs, with predictions frequently assigned to dominant or neighboring floe-size categories. Cake Ice is not detected effectively by any model, with per-class F1-scores remaining approximately zero across all configurations. These results indicate that FLOE prediction remains strongly affected by class imbalance, ambiguity among neighboring floe-size categories, and mixed conditions within the polygon-based ice-chart labels.

Overall, the per-class results highlight substantial class imbalance in Arctic-SAR, with performance degrading for poorly represented SOD and FLOE categories. This indicates that the main challenge is not only cross-sensor fusion, but also learning robust representations for under-represented ice classes.

\begin{table*}[!t]
\centering
\renewcommand{\arraystretch}{1.2}
\caption{Per-class FLOE F1-scores on the test set for the evaluated single-sensor baselines and cross-sensor fusion strategies. Values are reported as the mean $\pm$ standard deviation across five folds, with results averaged over three independent runs per fold.}
\label{tab:floe_per_class_f1}
\footnotesize
\begin{tabular}{lccccccc}
\hline
\textbf{Model}
& \textbf{Open Water [\%]}
& \textbf{Cake Ice [\%]}
& \textbf{Small Floe [\%]}
& \textbf{Medium Floe [\%]}
& \textbf{Big Floe [\%]}
& \textbf{Vast Floe [\%]}
& \textbf{Bergs [\%]} \\
\hline

Sentinel-1-only
& 97.39 $\pm$ 0.41
& 0.00 $\pm$ 0.00
& \textbf{19.69 $\pm$ 6.29}
& 7.63 $\pm$ 5.63
& 32.08 $\pm$ 9.93
& 76.15 $\pm$ 6.59
& \textbf{8.68 $\pm$ 2.45} \\

RCM-only
& 96.10 $\pm$ 0.85
& 0.00 $\pm$ 0.00
& 10.84 $\pm$ 7.35
& 3.93 $\pm$ 4.49
& 30.02 $\pm$ 4.81
& 75.66 $\pm$ 9.63
& 1.73 $\pm$ 1.99 \\

Input-level fusion
& 97.29 $\pm$ 0.41
& 0.00 $\pm$ 0.00
& 10.23 $\pm$ 3.71
& \textbf{10.62 $\pm$ 6.84}
& \textbf{32.69 $\pm$ 10.27}
& 77.70 $\pm$ 3.86
& 5.69 $\pm$ 4.21 \\

Feature-level fusion
& \textbf{97.42 $\pm$ 0.76}
& \textbf{0.01 $\pm$ 0.02}
& 13.13 $\pm$ 5.17
& 6.80 $\pm$ 2.63
& 31.39 $\pm$ 6.28
& 78.72 $\pm$ 7.09
& 6.75 $\pm$ 9.96 \\

Decision-level fusion
& 97.28 $\pm$ 0.44
& 0.00 $\pm$ 0.00
& 16.57 $\pm$ 7.06
& 4.55 $\pm$ 4.52
& 32.33 $\pm$ 7.96
& \textbf{78.73 $\pm$ 6.77}
& 4.97 $\pm$ 2.79 \\
\hline
\end{tabular}
\end{table*}

\subsection{Sensor Contribution and Fusion Implications}

The single-sensor results indicate that Sentinel-1 provides the stronger overall baseline in this benchmark, achieving a higher combined score than RCM on both the validation and test sets. However, the task- and class-specific results show that RCM may still provide useful information for selected prediction targets when combined with Sentinel-1. For example, input-level fusion achieves the highest SIC aAcc and $R^2$, feature-level fusion achieves the highest SOD aAcc and F1-score, and decision-level fusion achieves the highest FLOE aAcc and F1-score on the test set. Nevertheless, these improvements over Sentinel-1-only remain marginal, with input-level fusion achieving the highest combined score of $85.81\pm2.21\%$, compared with $85.48\pm2.00\%$ for Sentinel-1-only. The relative contribution of the two sensors is therefore dependent on the task, metric, and sea ice category, rather than consistently benefiting all prediction targets.

The cross-sensor backscatter intensity and local backscatter variability analyses in Fig.~\ref{fig:cross_sensor_scatter} provide additional context for this behavior. The co-registered Sentinel-1 and RCM backscatter values show moderate-to-strong correlations, indicating substantial shared C-band scattering information, while the systematic offsets and scatter demonstrate sensor-specific radiometric and spatial characteristics. The local standard deviation measures are also positively correlated for HH and HV. However, most points lie above the one-to-one line, indicating systematically stronger local backscatter variability in RCM. Therefore, the two sensors observe related scattering patterns but are not radiometrically identical.

These results also provide insights into the performance of different fusion strategies. Input-level fusion directly concatenates heterogeneous SAR channels with different radiometric distributions and noise characteristics, yet achieves the highest test combined score, indicating that direct channel concatenation can achieve competitive performance, although the additional benefit over Sentinel-1-only remains limited. Feature-level fusion allows each sensor branch to learn sensor-specific representations before multi-scale integration and achieves a comparable combined score, indicating that separate feature extraction does not provide a substantial advantage over direct input concatenation in this benchmark. Decision-level fusion preserves independently trained sensor-specific models and combines their outputs using task-specific validation-selected weights. Although it achieves the highest validation combined score, its test performance remains comparable to Sentinel-1-only, suggesting that validation-optimized sensor weighting does not necessarily translate into improved generalization. Overall, the marginal differences among fusion strategies suggest that the substantial shared information between Sentinel-1 and RCM limits the benefits of cross-sensor integration. Future cross-sensor SAR models could therefore focus on identifying complementary sensor information and adaptively integrating it, rather than simply increasing the amount of input data.

\subsection{Limitations and Future Work}
\label{subsec:limitations_future}
Despite the usefulness of the proposed cross-sensor SAR dataset, co-registration pipeline, and fusion benchmark, several limitations remain.

From a data perspective, Arctic-SAR is constrained by the availability of temporally close Sentinel-1 and RCM acquisitions with sufficient spatial overlap. This limits the dataset size and contributes to class imbalance, where dominant classes occur much more frequently than rare SIC, SOD, and FLOE categories. Accordingly, the challenge in Arctic-SAR extends beyond cross-sensor fusion itself to learning robust representations for under-represented ice classes. Although geographic co-registration aligns the two SAR products spatially, temporal separation can still introduce residual mismatch because sea ice may move or deform between acquisitions. Since 95\% of the paired acquisitions are separated by less than approximately 48~min, temporal mismatch is limited for most pairs, although the few scenes with larger acquisition offsets may be more affected by sea ice motion. Furthermore, the mean co-registration RMSE of 6.60 pixels (approximately 264~m) indicates residual spatial misalignment, which may limit the potential benefits of cross-sensor fusion by reducing the spatial correspondence between the two sensor observations.

In addition, the reference labels are derived from operational ice charts, which provide polygon-level annotations rather than true pixel-level labels. As a result, mixed ice conditions, small-scale spatial variability, and uncertainty in intermediate SIC levels, transitional SOD classes, and minority FLOE classes may not be fully represented. Moreover, the RCM data have not yet been denoised using a noise-floor removal procedure comparable to that applied to Sentinel-1, which may leave residual noise, sub-swath boundary effects, and sensor-related artifacts. Increasing the number of co-registered scenes, constructing a more balanced dataset, and applying RCM-specific denoising would likely improve model robustness and rare-class performance. 

From a modeling perspective, the current fusion framework is limited by its CNN-based U-Net backbone and relatively simple convolutional feature-fusion blocks. Although the evaluated fusion strategies achieve competitive performance, five-fold cross-validation shows that their improvements over the Sentinel-1-only baseline remain marginal. The current feature-level design mainly integrates Sentinel-1 and RCM representations through local convolutional operations and therefore may not fully capture long-range spatial dependencies, large-scale sea ice structures, or complex cross-sensor interactions. This limitation is particularly relevant for heterogeneous SAR fusion because Sentinel-1 and RCM differ in radiometric distribution, calibration properties, acquisition geometry, and noise characteristics. Therefore, the most informative sensor contribution may vary with task, ice class, spatial location, and local imaging conditions. However, the current convolutional fusion strategy does not explicitly learn where, when, or how much each sensor should contribute. More adaptive cross-sensor fusion mechanisms may therefore be required to fully exploit the complementary information from both platforms more consistently across unseen scenes. 

The proposed Arctic-SAR dataset provides a basis for future research on attention-guided feature-level fusion, particularly transformer-based and cross-attention designs. Transformer-based fusion can provide broader spatial context through self-attention, while cross-attention can explicitly model interactions between the two sensor streams by allowing features from one sensor to attend to complementary features from the other~\cite{Roy2023MFT, Ge2024CAFM}. This may be particularly useful for heterogeneous SAR fusion, where Sentinel-1 and RCM differ in radiometry, noise level, acquisition geometry, and class-specific reliability. Such approaches could help identify sensor-specific complementary information and adaptively integrate useful features across spatial scales, tasks, and ice conditions.

\section{Conclusion}
\label{sec:conclusion}
This study introduced Arctic-SAR, a co-registered cross-sensor SAR dataset that aligns RADARSAT Constellation Mission (RCM) dual-polarization imagery with Sentinel-1 scenes from the AI4Arctic Sea Ice Dataset for automated prediction of sea ice concentration (SIC), stage of development (SOD), and floe size (FLOE) using ice-chart-derived reference labels. In addition to the dataset, this work established a systematic benchmark for evaluating cross-sensor SAR fusion through five configurations: Sentinel-1-only, RCM-only, input-level fusion, feature-level fusion, and decision-level fusion.

The experimental results reveal task- and class-dependent performance differences between Sentinel-1 and RCM. Overall, Sentinel-1 provides a stronger single-sensor baseline than RCM. Furthermore, the evaluated fusion strategies provide only marginal improvements over the stronger single-sensor baseline. Among the fusion strategies, input-level fusion achieves the highest combined score, improving performance by only 0.33 percentage points over Sentinel-1 alone. These findings suggest that RCM provides limited additional predictive information beyond Sentinel-1, as none of the evaluated fusion strategies consistently improves performance across all three sea-ice mapping tasks.

Arctic-SAR provides a new benchmark for studying cross-sensor SAR fusion in operational sea ice mapping and highlights the challenges and limited additional predictive benefits of incorporating RCM observations with Sentinel-1 imagery. Future work will explore sensor-aware and attention-based fusion strategies that can preserve sensor-specific representations and dynamically integrate complementary information across tasks and sea ice conditions.

\section*{Acknowledgments}
The authors acknowledge the Canadian Space Agency for access to RCM data and the Canadian Ice Service for providing the operational sea ice charts used in the dataset. We acknowledge the support of the Natural Sciences and Engineering Research Council of Canada (NSERC) under Grant RGPIN-2024-04020 and the Canadian Space Agency (CSA) through the Research Opportunities in Space Science program under Grant 23SUESDEEP.


\begin{thebibliography}{99}
\bibliographystyle{IEEEtran}

\bibitem{Chen2024}
X. Chen, M. Patel, F. J. Pena Cantu, J. Park, J. Noa Turnes, L. Xu, K. A. Scott, and D. A. Clausi, ``MMSeaIce: A collection of techniques for improving sea ice mapping with a multi-task model,'' \textit{The Cryosphere}, vol. 18, pp. 1621--1632, 2024. doi: 10.5194/tc-18-1621-2024.

\bibitem{Lyu2022}
H. Lyu, W. Huang, and M. Mahdianpari,
``A meta-analysis of sea ice monitoring using spaceborne polarimetric SAR: Advances in the last decade,''
\textit{IEEE J. Sel. Topics Appl. Earth Observ. Remote Sens.},
vol. 15, pp. 6158--6179, 2022,
doi: 10.1109/JSTARS.2022.3194324.

\bibitem{Swart2015}
N. C. Swart, J. C. Fyfe, E. Hawkins, J. E. Kay, and A. Jahn,
``Influence of internal variability on Arctic sea-ice trends,''
\textit{Nature Climate Change},
vol. 5, no. 2, pp. 86--89, 2015,
doi: 10.1038/nclimate2483.

\bibitem{Ye2023}
Y. Ye, Y. Luo, Y. Sun, M. Shokr, S. Aaboe, F. Girard-Ardhuin,
F. Hui, X. Cheng, and Z. Chen,
``Inter-comparison and evaluation of Arctic sea ice type products,''
\textit{The Cryosphere}, vol. 17, no. 1, pp. 279--308, 2023.
doi: 10.5194/tc-17-279-2023.


\bibitem{Ma2026GLFFuse}
T. Ma, X. Chen, H. Qin, L. Xu, P. Yu, and W. Huang, ``GLFFuse: A multimodal feature-level fusion network for multitask fine-grained recognition of Arctic sea ice,'' \textit{IEEE J. Sel. Topics Appl. Earth Observ. Remote Sens.}, vol. 19, pp. 7661--7679, 2026. doi: 10.1109/JSTARS.2026.3660828.


\bibitem{Zakhvatkina2019}
N. Zakhvatkina, V. Smirnov, and I. Bychkova, ``Satellite SAR data-based sea ice classification: An overview,'' \textit{Geosciences}, vol. 9, no. 4, Art. no. 152, 2019. doi: 10.3390/geosciences9040152.


\bibitem{Turnes2025ITT}
J. N. Turnes, M. Jiang, S. Taleghanidoozdoozan, L. Xu, and D. A. Clausi, ``ITT: Long-range spatial dependencies for sea ice semantic segmentation,'' \textit{IEEE J. Sel. Topics Appl. Earth Observ. Remote Sens.}, vol. 18, pp. 13296--13309, 2025. doi: 10.1109/JSTARS.2025.3570896.

\bibitem{Dierking2013}
W. Dierking, ``Sea ice monitoring by synthetic aperture radar,'' \textit{Oceanography}, vol. 26, no. 2, pp. 100--111, 2013. doi: 10.5670/oceanog.2013.33.

\bibitem{Soh1999}
L.-K. Soh and C. Tsatsoulis, ``Texture analysis of SAR sea ice imagery using gray level co-occurrence matrices,'' \textit{IEEE Trans. Geosci. Remote Sens.}, vol. 37, no. 2, pp. 780--795, Mar. 1999. doi: 10.1109/36.752194.

\bibitem{Boulze2020}
H. Boulze, A. Korosov, and J. Brajard, ``Classification of sea ice types in Sentinel-1 SAR data using convolutional neural networks,'' \textit{Remote Sensing}, vol. 12, no. 13, Art. no. 2165, 2020. doi: 10.3390/rs12132165.

\bibitem{Han2021}
Y. Han, Y. Liu, Z. Hong, Y. Zhang, S. Yang, and J. Wang, ``Sea ice image classification based on heterogeneous data fusion and deep learning,'' \textit{Remote Sensing}, vol. 13, no. 4, Art. no. 592, 2021. doi: 10.3390/rs13040592.

\bibitem{Leigh2014}
S. Leigh, Z. Wang, and D. A. Clausi, ``Automated ice-water classification using dual polarization SAR satellite imagery,'' \textit{IEEE Trans. Geosci. Remote Sens.}, vol. 52, no. 9, pp. 5529--5539, Sep. 2014. doi: 10.1109/TGRS.2013.2290231.

\bibitem{Gerland2019}
S. Gerland, D. Barber, W. Meier, C. J. Mundy, M. Holland, S. Kern,
Z. Li, C. Michel, D. K. Perovich, and T. Tamura,
``Essential gaps and uncertainties in the understanding of the roles and
functions of Arctic sea ice,'' \textit{Environmental Research Letters},
vol. 14, no. 4, Art. no. 043002, Apr. 2019.
doi: 10.1088/1748-9326/ab09b3.

\bibitem{WangLi2021}
Y. Wang and X. Li, ``Arctic sea ice cover data from spaceborne synthetic aperture radar by deep learning,'' \textit{Earth Syst. Sci. Data}, vol. 13, no. 6, pp. 2723--2742, Jun. 2021. doi: 10.5194/essd-13-2723-2021.

\bibitem{deLoe2025}
L. de Loe, D. A. Clausi, and K. A. Scott,
``Fusing ice surface temperature with the AI4Arctic dataset for improved deep learning-based sea ice mapping,''
\textit{IEEE J. Sel. Topics Appl. Earth Observ. Remote Sens.},
vol. 18, pp. 24766--24782, 2025,
doi: 10.1109/JSTARS.2025.3610260.

\bibitem{Kruk2020}
R. Kruk, M. C. Fuller, A. S. Komarov, D. Isleifson, and I. Jeffrey, ``Proof of concept for sea ice stage of development classification using deep learning,'' \textit{Remote Sensing}, vol. 12, no. 15, Art. no. 2486, 2020. doi: 10.3390/rs12152486.

\bibitem{Khaleghian2021}
S. Khaleghian, H. Ullah, T. Kr{\ae}mer, N. Hughes, T. Eltoft, and A. Marinoni, ``Sea ice classification of SAR imagery based on convolution neural networks,'' \textit{Remote Sensing}, vol. 13, no. 9, Art. no. 1734, 2021. doi: 10.3390/rs13091734.

\bibitem{Zakhvatkina2013}
N. Y. Zakhvatkina, V. Y. Alexandrov, O. M. Johannessen, S. Sandven, and I. Y. Frolov, ``Classification of sea ice types in ENVISAT synthetic aperture radar images,'' \textit{IEEE Trans. Geosci. Remote Sens.}, vol. 51, no. 5, pp. 2587--2600, May 2013. doi: 10.1109/TGRS.2012.2212445.

\bibitem{Ressel2016}
R. Ressel, S. Singha, S. Lehner, A. R{\"o}sel, and G. Spreen,
``Investigation into different polarimetric features for sea ice classification using X-band synthetic aperture radar,''
\textit{IEEE J. Sel. Topics Appl. Earth Observ. Remote Sens.},
vol. 9, no. 7, pp. 3131--3143, Jul. 2016,
doi: 10.1109/JSTARS.2016.2539501.

\bibitem{Deng2005}
H. Deng and D. A. Clausi, ``Unsupervised segmentation of synthetic aperture radar sea ice imagery using a novel Markov random field model,'' \textit{IEEE Trans. Geosci. Remote Sens.}, vol. 43, no. 3, pp. 528--538, Mar. 2005. doi: 10.1109/TGRS.2004.839589.

\bibitem{Rogers2024}
M. S. J. Rogers, M. Fox, A. Fleming, L. van Zeeland, J. Wilkinson, and J. S. Hosking, ``Sea ice detection using concurrent multi-spectral and synthetic aperture radar imagery,'' \textit{Remote Sens. Environ.}, vol. 305, Art. no. 114073, 2024. doi: 10.1016/j.rse.2024.114073.


\bibitem{Radhakrishnan2021}
K. Radhakrishnan, K. A. Scott, and D. A. Clausi, ``Sea ice concentration estimation: Using passive microwave and SAR data with a U-Net and curriculum learning,'' \textit{IEEE J. Sel. Topics Appl. Earth Observ. Remote Sens.}, vol. 14, pp. 5339--5351, 2021. doi: 10.1109/JSTARS.2021.3076109.


\bibitem{Stokholm2022}
A. Stokholm, T. Wulf, A. Kucik, R. Saldo, J. Buus-Hinkler, and S. M. Hvidegaard, ``AI4SeaIce: Towards solving ambiguous SAR textures in convolutional neural networks for automatic sea ice concentration charting,'' \textit{IEEE Trans. Geosci. Remote Sens.}, vol. 60, Art. no. 4304013, pp. 1--13, 2022. doi: 10.1109/TGRS.2022.3149323.


\bibitem{Jamali2024}
A. Jamali, S. K. Roy, L. Hashemi Beni, B. Pradhan, J. Li, and P. Ghamisi, ``Residual wave vision U-Net for flood mapping using dual polarization Sentinel-1 SAR imagery,'' \textit{Int. J. Appl. Earth Observ. Geoinf.}, vol. 127, Art. no. 103662, Mar. 2024. doi: 10.1016/j.jag.2024.103662.

\bibitem{Bonafilia2020}
D. Bonafilia, B. Tellman, T. Anderson, and E. Issenberg,
``Sen1Floods11: A georeferenced dataset to train and test deep learning flood algorithms for Sentinel-1,''
in \textit{Proc. IEEE/CVF Conf. Comput. Vis. Pattern Recognit. Workshops},
Jun. 2020, pp. 835--845,
doi: 10.1109/CVPRW50498.2020.00113.

\bibitem{Ghamisi2019}
P. Ghamisi, B. Rasti, N. Yokoya, Q. Wang, B. H{\"o}fle, L. Bruzzone, F. Bovolo, M. Chi, K. Anders, R. Gloaguen, P. M. Atkinson, and J. A. Benediktsson,
``Multisource and multitemporal data fusion in remote sensing: A comprehensive review of the state of the art,''
\textit{IEEE Geosci. Remote Sens. Mag.},
vol. 7, no. 1, pp. 6--39, Mar. 2019.
doi: 10.1109/MGRS.2018.2890023.

\bibitem{Khachatrian2023}
E. Khachatrian, W. Dierking, S. Chlaily, T. Eltoft, F. Dinessen, N. Hughes, and A. Marinoni, ``SAR and passive microwave fusion scheme: A test case on Sentinel-1/AMSR-2 for sea ice classification,'' \textit{Geophys. Res. Lett.}, vol. 50, no. 4, Art. no. e2022GL102083, Feb. 2023. doi: 10.1029/2022GL102083.

\bibitem{Gao2025}
T. Gao, C. Lan, C. Zhou, Y. Zhang, W. Huang, Y. Wang, and L. Wang,
``Arctic sea ice motion retrieval from multisource SAR images using a
keypoint-free feature tracking algorithm,''
\textit{ISPRS J. Photogramm. Remote Sens.},
vol. 230, pp. 258--274, 2025.
doi: 10.1016/j.isprsjprs.2025.09.013.


\bibitem{Johansson2020}
A. M. Johansson, E. Malnes, S. Gerland, A. Cristea,
A. P. Doulgeris, D. V. Divine, O. Pavlova, and T. R. Lauknes,
``Consistent ice and open water classification combining historical
synthetic aperture radar satellite images from ERS-1/2, Envisat ASAR,
RADARSAT-2 and Sentinel-1A/B,''
\textit{Annals of Glaciology}, vol. 61, no. 82, pp. 40--50, 2020.
doi: 10.1017/aog.2019.52.

\bibitem{Howell2018}
S. E. L. Howell, A. S. Komarov, M. Dabboor, B. Montpetit, M. Brady,
R. K. Scharien, M. S. Mahmud, V. Nandan, T. Geldsetzer, and J. J. Yackel,
``Comparing L- and C-band synthetic aperture radar estimates of sea ice
motion over different ice regimes,''
\textit{Remote Sens. Environ.}, vol. 204, pp. 380--391, 2018.
doi: 10.1016/j.rse.2017.10.017.

\bibitem{Marsocci2025}
V. Marsocci and N. Audebert, ``Cross-sensor self-supervised training and alignment for remote sensing,'' \textit{IEEE J. Sel. Topics Appl. Earth Observ. Remote Sens.}, vol. 18, pp. 12278--12289, 2025. doi: 10.1109/JSTARS.2025.3566457.


\bibitem{BuusHinkler2022}
J. Buus-Hinkler, T. Wulf, A. R. Stokholm, A. Korosov, R. Saldo,
L. T. Pedersen, D. Arthurs, R. Solberg, N. Long{\'e}p{\'e}, and M. B. Kreiner,
``AI4Arctic Sea Ice Challenge Dataset,'' Technical University of Denmark,
2022. doi: 10.11583/DTU.c.6244065.v2.

\bibitem{Kortum2022}
K. Kortum, S. Singha, and G. Spreen, ``Robust multiseasonal ice classification from high-resolution X-band SAR,'' \textit{IEEE Trans. Geosci. Remote Sens.}, vol. 60, Art. no. 4408512, pp. 1--12, 2022. doi: 10.1109/TGRS.2022.3144731.

\bibitem{Ronneberger2015}
O. Ronneberger, P. Fischer, and T. Brox, ``U-Net: Convolutional networks for biomedical image segmentation,'' in \textit{Proc. Medical Image Computing and Computer-Assisted Intervention (MICCAI)}, Munich, Germany, 2015, pp. 234--241. doi: 10.1007/978-3-319-24574-4-28.


\bibitem{Huang2024}
Y. Huang, Y. Ren, and X. Li, ``Deep learning techniques for enhanced sea-ice types classification in the Beaufort Sea via SAR imagery,'' \textit{Remote Sens. Environ.}, vol. 308, Art. no. 114204, Jul. 2024. doi: 10.1016/j.rse.2024.114204.

\bibitem{Ren2021}
Y. Ren, X. Li, X. Yang, and H. Xu, ``Development of a dual-attention U-Net model for sea ice and open water classification on SAR images,'' \textit{IEEE Geosci. Remote Sens. Lett.}, vol. 19, Art. no. 4010205, pp. 1--5, 2022. doi: 10.1109/LGRS.2021.3058049.

\bibitem{ESA_Sentinel1}
European Space Agency, ``Sentinel-1: Radar vision for Copernicus,'' ESA Earth Observation. [Online]. Available: \url{https://www.esa.int/Applications/Observing_the_Earth/Copernicus/Sentinel-1}. 

\bibitem{Xu2018}
X. Xu, W. Li, Q. Ran, Q. Du, L. Gao, and B. Zhang, ``Multisource remote sensing data classification based on convolutional neural network,'' \textit{IEEE Trans. Geosci. Remote Sens.}, vol. 56, no. 2, pp. 937--949, Feb. 2018. doi: 10.1109/TGRS.2017.2756851.

\bibitem{Ma2024}
X. Ma, X. Zhang, M.-O. Pun, and M. Liu,
``A multilevel multimodal fusion transformer for remote sensing semantic segmentation,''
\textit{IEEE Trans. Geosci. Remote Sens.},
vol. 62, Art. no. 5403215, pp. 1--15, 2024,
doi: 10.1109/TGRS.2024.3373033.

\bibitem{Pereira2024}
L. M. Pereira, A. Salazar, and L. Vergara, ``A comparative study on recent automatic data fusion methods,'' \textit{Computers}, vol. 13, no. 1, Art. no. 13, 2024. doi: 10.3390/computers13010013.

\bibitem{Huang2019}
J. Huang, X. Zhang, Q. Xin, Y. Sun, and P. Zhang, ``Automatic building extraction from high-resolution aerial images and LiDAR data using gated residual refinement network,'' \textit{ISPRS J. Photogramm. Remote Sens.}, vol. 151, pp. 91--105, May 2019. doi: 10.1016/j.isprsjprs.2019.02.019.

\bibitem{Wei2024MGFNet}
K. Wei, J. Dai, D. Hong, and Y. Ye, ``MGFNet: An MLP-dominated gated fusion network for semantic segmentation of high-resolution multimodal remote sensing images,'' \textit{Int. J. Appl. Earth Obs. Geoinf.}, vol. 135, Art. no. 104241, 2024. doi: 10.1016/j.jag.2024.104241.

\bibitem{Liu2024SoftFormer}
R. Liu, J. Ling, and H. Zhang,
``SoftFormer: SAR--optical fusion transformer for urban land use and land cover classification,''
\textit{ISPRS J. Photogramm. Remote Sens.},
vol. 218, pp. 277--293, 2024,
doi: 10.1016/j.isprsjprs.2024.09.012.


\bibitem{Xiong2023}
Z. Xiong, S. Chen, Y. Wang, L. Mou, and X. X. Zhu,
``GAMUS: A geometry-aware multi-modal semantic segmentation benchmark for remote sensing data,''
\textit{arXiv preprint arXiv:2305.14914}, 2023,
doi: 10.48550/arXiv.2305.14914.

\bibitem{WangDecisionFusion}
L. Wang, K. A. Scott, and D. A. Clausi, ``Improved sea ice concentration estimation through fusing classified SAR imagery and AMSR-E data,'' \textit{Can. J. Remote Sens.}, vol. 42, no. 1, pp. 41--52, 2016. doi: 10.1080/07038992.2016.1152547.

\bibitem{Korosov2022}
A. Korosov, D. Demchev, N. Miranda, N. Franceschi, and J.-W. Park,
``Thermal denoising of cross-polarized Sentinel-1 data in interferometric and extra wide swath modes,''
\textit{IEEE Trans. Geosci. Remote Sens.},
vol. 60, Art. no. 5218411, pp. 1--11, 2022,
doi: 10.1109/TGRS.2021.3131036.

\bibitem{manice2005}
Environment Canada, ``Manual of Standard Procedures for Observing and
Reporting Ice Conditions (MANICE),'' Meteorological Service of Canada,
2005. [Online]. Available:
\url{https://www.canada.ca/en/environment-climate-change/services/weather-manuals-documentation/manice-manual-of-ice.html}.


\bibitem{Stokholm2024}
A. Stokholm \textit{et al.}, ``The AutoICE Challenge,'' \textit{The Cryosphere}, vol. 18, pp. 3471--3494, 2024. doi: 10.5194/tc-18-3471-2024.

\bibitem{EODMS} Natural Resources Canada, ``Earth Observation Data Management System (EODMS),'' Government of Canada. [Online]. Available: \url{https://www.eodms-sgdot.nrcan-rncan.gc.ca/index-en.html}.

\bibitem{CMEMSSeaIceDriftData}
Copernicus Marine Service,
``Global Ocean -- High Resolution SAR Sea Ice Drift Time Series,''
Product ID: SEAICE\_GLO\_PHY\_L4\_MY\_011\_020.
doi: 10.48670/mds-00328.
[Online]. Available:
\url{https://data.marine.copernicus.eu/product/SEAICE_GLO_PHY_L4_MY_011_020/description}

\bibitem{CMEMSSeaIceDrift}
R. Saldo,
``Global Ocean -- High Resolution SAR Sea Ice Drift: Product User Manual,''
Copernicus Marine Service, CMEMS-SEAICE-PUM-011-006-020, 2023.
[Online]. Available:
\url{https://documentation.marine.copernicus.eu/PUM/CMEMS-SEAICE-PUM-011-006-020.pdf}

\bibitem{QiuLi2026}
Y. Qiu and X.-M. Li,
``Pan-Arctic sea ice drift data at 400 m grids based on spaceborne SAR,''
\textit{Scientific Data}, vol. 13, no. 1, Art. no. 377, 2026.
doi: 10.1038/s41597-026-06683-9.

\bibitem{RCMProductFormat}
MDA, ``RADARSAT Constellation Mission Image Product Format Definition,''
RCM-SP-53-0419, Issue 2/8/P, June 2021. [Online]. Available:
\url{https://download-telecharger.services.geo.ca/pub/csa_asc/Space-technology_Technologie-spatiale/radarsat_constellation_mission_plan/RCM-SP-53-0419_Image_Product_Format_Definition_2-8_Public.pdf}

\bibitem{Li2015}
J.-W. Li, Z.-F. Li, Y.-L. Hou, and Z. Bao,
``Efficient method for spaceborne synthetic aperture radar (SAR)
geolocation and application to geometrical SAR image registration,''
\textit{IET Radar, Sonar \& Navigation},
vol. 9, no. 1, pp. 67--74, 2015,
doi: 10.1049/iet-rsn.2013.0302.

\bibitem{Sansosti2006}
E. Sansosti, P. Berardino, M. Manunta, F. Serafino, and G. Fornaro,
``Geometrical SAR image registration,''
\textit{IEEE Trans. Geosci. Remote Sens.},
vol. 44, no. 10, pp. 2861--2870, Oct. 2006,
doi: 10.1109/TGRS.2006.875787.

\bibitem{Tang2024}
G. Tang, Z. Wei, and L. Zhuang,
``SAR Image Registration: The Combination of Nonlinear Diffusion Filtering,
Hessian Features and Edge Points,''
\textit{Sensors}, vol. 24, no. 14, Art. no. 4568, Jul. 2024,
doi: 10.3390/s24144568.

\bibitem{Unidata2020NetCDF}
Unidata, ``Network Common Data Form (NetCDF),'' UCAR/NSF Unidata Program Center, Boulder, CO, 2020. doi: 10.5065/D6H70CW6.

\bibitem{Jiang2022}
J. Jiang, K. Johansen, Y.-H. Tu, and M. F. McCabe,
``Multi-sensor and multi-platform consistency and interoperability between UAV, Planet CubeSat, Sentinel-2, and Landsat reflectance data,''
\textit{GIScience \& Remote Sensing}, vol. 59, no. 1, pp. 936--958, 2022.
doi: 10.1080/15481603.2022.2083791.

\bibitem{Samadzadegan2025}
F. Samadzadegan, A. Toosi, and F. Dadrass Javan,
``A critical review on multi-sensor and multi-platform remote sensing data fusion approaches: current status and prospects,''
\textit{Int. J. Remote Sens.}, vol. 46, no. 3, pp. 1327--1402, 2025.
doi: 10.1080/01431161.2024.2429784.

\bibitem{Zhang2026CrossSensor}
X. Zhang, S. Zhang, Z. Sun, C. Liu, Y. Sun, K. Ji, and G. Kuang,
``Semantic scattering graph structure alignment for cross-sensor SAR image target detection,''
\textit{Chinese Journal of Aeronautics}, vol. 39, no. 6, Art. no. 104066, 2026.
doi: 10.1016/j.cja.2026.104066.

\bibitem{Mouche2015Global}
A. Mouche and B. Chapron,
``Global C-band Envisat, RADARSAT-2 and Sentinel-1 SAR measurements in copolarization and cross-polarization,''
\textit{J. Geophys. Res. Oceans},
vol. 120, no. 11, pp. 7195--7207, 2015.
doi: 10.1002/2015JC011149.

\bibitem{Aiazzi2005Heterogeneity}
B. Aiazzi, L. Alparone, and S. Baronti,
``Information-theoretic heterogeneity measurement for SAR imagery,''
\textit{IEEE Trans. Geosci. Remote Sens.}, vol. 43, no. 3, pp. 619--624, 2005.
doi: 10.1109/TGRS.2004.837328.

\bibitem{Jiang2021}
Y. Jiang, H. Yao, S. Tao, and J. Liang, ``Gated skip-connection network with adaptive upsampling for retinal vessel segmentation,'' \textit{Sensors}, vol. 21, no. 18, Art. no. 6177, 2021. doi: 10.3390/s21186177.

\bibitem{Loshchilov2016}
I. Loshchilov and F. Hutter, ``SGDR: Stochastic gradient descent with warm restarts,'' \textit{arXiv preprint}, 2016. doi: 10.48550/arXiv.1608.03983.

\bibitem{Cheng2020}
A. Cheng, B. Casati, A. Tivy, T. Zagon, J.-F. Lemieux, and L. B. Tremblay,
``Accuracy and inter-analyst agreement of visually estimated sea ice concentrations in Canadian Ice Service ice charts using single-polarization RADARSAT-2,''
\textit{The Cryosphere}, vol. 14, no. 4, pp. 1289--1310, 2020.
doi: 10.5194/tc-14-1289-2020.


\bibitem{Roy2023MFT}
S. K. Roy, A. Deria, D. Hong, B. Rasti, A. Plaza, and J. Chanussot, 
``Multimodal fusion transformer for remote sensing image classification,'' 
\textit{IEEE Trans. Geosci. Remote Sens.}, vol. 61, pp. 1--20, 2023, Art. no. 5515620. doi: 10.1109/TGRS.2023.3286826.

\bibitem{Ge2024CAFM}
H. Ge, L. Wang, H. Pan, Y. Liu, C. Li, D. Lv, and H. Ma, 
``Cross attention-based multi-scale convolutional fusion network for hyperspectral and LiDAR joint classification,'' 
\textit{Remote Sensing}, vol. 16, no. 21, Art. no. 4073, 2024. doi: 10.3390/rs16214073.

\end{thebibliography}
\end{document}